\documentclass[aps,prb,preprint,superscriptaddress,reprint,preprintnumbers,amssymb]{revtex4-1}

\usepackage{graphicx}
\usepackage[latin1]{inputenc}
\usepackage{braket}
\usepackage{pifont}
\usepackage{units}
\usepackage{wrapfig}
\usepackage{graphicx}
\usepackage{amsmath}
\usepackage{booktabs}
\usepackage{rotating}
\usepackage{color}
\usepackage{amsfonts}
\usepackage{amssymb}
\usepackage{moreverb}
\usepackage{url}
\usepackage{epsf}
\usepackage{multirow}
\usepackage{subfigure}
\usepackage{morefloats}
\usepackage{threeparttable}
\usepackage{cancel}
\usepackage{bm}
\usepackage{tensor}
\usepackage[version=3]{mhchem}

\DeclareMathOperator{\tr}{tr}

\begin{document}

\title{Accurate and efficient linear scaling DFT calculations with universal applicability}

\author{Stephan Mohr}
%\email{}
\affiliation{Univ.\ Grenoble Alpes, CEA, INAC-SP2M, F-38000 Grenoble, France }

\author{Laura E.\ Ratcliff}
\affiliation{Univ.\ Grenoble Alpes, CEA, INAC-SP2M, F-38000 Grenoble, France }
\affiliation{Argonne Leadership Computing Facility, Argonne National Laboratory, Illinois 60439, USA}

\author{Luigi Genovese}
\email{luigi.genovese@cea.fr}
\affiliation{Univ.\ Grenoble Alpes, CEA, INAC-SP2M, F-38000 Grenoble, France }

\author{Damien Caliste}
%\email{}
\affiliation{Univ.\ Grenoble Alpes, CEA, INAC-SP2M, F-38000 Grenoble, France }

\author{Paul Boulanger}
\affiliation{Univ.\ Grenoble Alpes, CEA, INAC-SP2M, F-38000 Grenoble, France }

\author{Stefan Goedecker}
\affiliation{Institut f\"ur Physik, Universit\"at Basel, Klingelbergstr.\ 82, 4056 Basel, Switzerland}

\author{Thierry Deutsch}
%\email{}
\affiliation{Univ.\ Grenoble Alpes, CEA, INAC-SP2M, F-38000 Grenoble, France }

\date{\today}

\begin{abstract}
%# OLD ##############################################
% Density Functional Theory (DFT) is one of the most powerful ab-initio methods for performing electronic structure calculations thanks to its good balance between accuracy and speed. Consequently it is widely used in various fields such as physics, chemistry, biology and material sciences. Thanks to the steadily increasing power of nowadays supercomputers, the size of the systems which can be investigated could be continuously increased over the years. However classical DFT suffers from an inherent cubic scaling with respect to the size of the system, making big calculations extremely expensive. Fortunately this cubic scaling can be avoided by the use of so-called linear scaling algorithms, which have been developed during the last few decades. In this way it becomes possible to perform ab-initio calculations of several ten thousands atoms or even more within a reasonable time frame. Even though the use of such linear scaling algorithms is physically well justified, it often still introduces some errors. 
% However, we were able to implement a fully linear scaling version of the well established BigDFT code which exhibits an amazingly high accuracy and a universal applicability while still allowing to simulate huge systems with an only moderate demand of computing resources.
%# END OLD ##########################################
%# NEW ##############################################
Density Functional Theory calculations traditionally suffer from an inherent cubic scaling with respect to the size of the system, making big calculations extremely expensive. This cubic scaling can be avoided by the use of so-called linear scaling algorithms, which have been developed during the last few decades. In this way it becomes possible to perform ab-initio calculations for several tens of thousands of atoms or even more within a reasonable time frame. However, even though the use of linear scaling algorithms is physically well justified, their implementation often introduces some small errors. Consequently most implementations offering such a linear complexity either yield only a limited accuracy or, if one wants to go beyond this restriction, require a tedious fine tuning of many parameters. In our linear scaling approach within the BigDFT package, we were able to overcome this restriction. Using an ansatz based on localized support functions expressed in an underlying Daubechies wavelet basis 
-- which offers ideal properties for accurate linear scaling calculations -- we obtain an amazingly high accuracy and a universal applicability while still keeping the possibility of simulating large systems with only a moderate demand of computing resources.
%virtually the same precision as for the traditional cubic scaling approach using a standard set of parameters. Moreover the 
%# END NEW ##########################################
\end{abstract}

%\pacs
%{
%}

\maketitle

\section{Introduction} 
The Kohn-Sham formalism of Density Functional Theory (DFT)~\cite{hohenberg1964inhomogeneous,kohn1965celf-consistent} has established itself as one of the most powerful electronic structure methods due to its good balance between accuracy and speed and is thus popular in various fields such as physics, chemistry, biology, and material sciences. Kohn-Sham DFT maps the problem of interacting electrons onto a problem of non-interacting quasi-electrons: given a system containing $N$ electrons, the $3N$-dimensional many-electron wave function is assumed to be given by a single Slater determinant built from $N$ orthonormal single particle orbitals $\psi(\mathbf{r})$ which a priori extend over the entire system. The drawback of this approach is that it has an inherent cubic scaling due to the orthonormality which is imposed on these so-called Kohn-Sham orbitals. To orthonormalize a set of $N$ functions it is necessary to calculate the scalar products among all of them, which has the complexity $\mathcal{O}(N^2)$. In addition the 
cost of calculating one scalar product is proportional to the size of the underlying basis, $m_{basis}$. Since both $N$ and $m_{basis}$ are proportional to the total size of the system -- 
typically indicated by the number of atoms $n_{at}$ -- one ends up with a complexity $\mathcal{O}(n_{at}^3)$. This behavior is common to all programs using a systematic basis set, be it plane waves~\cite{gonze2009abinit,giannozzi2009quantum,segall2002first-principles}, finite elements~\cite{pask2005finite} or wavelets~\cite{genovese2008daubechies}. Even when not using a systematic basis set, e.g.\ in the case of Gaussians~\cite{valiev2010NWChem} or atomic orbitals~\cite{blum2009ab}, the complexity remains $\mathcal{O}(n_{at}^3)$
%, therefore cubically scaling with respect to $N$, 
due to the matrix diagonalization which is required in a straightforward implementation. Due to this limitation this standard approach is only suitable for systems containing a few hundred or at most a few thousand atoms.

In order to obtain an algorithm which can scale linearly with respect to the size of the system, one thus has to abandon the concept of the extended Kohn-Sham orbitals and work with other quantities which are strictly localized.
% To this end one can express the Kohn-Sham orbitals in terms of a set of strictly localized support functions, $\psi_i(\mathbf{r})=\sum_\alpha c_i^\alpha\phi^\alpha(\mathbf{r})$.
One such quantity is the density matrix $\mathbf{F}(\mathbf{r},\mathbf{r}')$, which is related to the Kohn-Sham orbitals via $\mathbf{F}(\mathbf{r},\mathbf{r}') = \sum_i f_i\psi_i(\mathbf{r})\psi_i(\mathbf{r}')$ with $f_i$ being the occupation number of orbital $i$. For insulators or metals at finite temperature it can be shown that the elements of $\mathbf{F(\mathbf{r},\mathbf{r}')}$ decay exponentially with the distance between $\mathbf{r}$ and $\mathbf{r}'$~\cite{cloizeaux1964energy,cloizeaux1964analytical,kohn1959analytic,baer1997sparsity,ismail-beigi1999locality,goedecker1998decay,he2001exponential}. Consequently, if one neglects elements which are below a given threshold, the number of non-zero elements scales only linearly with respect to the size of the system, thus paving the way towards an algorithm with this reduced complexity. During the past decades this fact has led to a number of algorithms which are capable of performing linear scaling calculations; an overview is given in Ref.~\onlinecite{goedecker1999linear}.

One particular approach is to write the density matrix in separable form by introducing a set of so-called support functions~\cite{hierse1994order,hernandez1995self-consistent}; an idea that has already been used in various codes such as ONETEP~\cite{skylaris2005introducing}, Conquest~\cite{bowler2010calculations}, CP2K~\cite{vandevondele2005quickstep} and SIESTA~\cite{soler2002the}. These support functions can also be thought of as a localized basis to directly represent the Kohn-Sham orbitals. Even if there exists an ideal set of localized functions -- namely the maximally localized Wannier functions~\cite{marzari1997maximally} -- this is of course not known beforehand. 

%We have recently implemented a linear scaling code based on Daubechies wavelets, distributed in the BigDFT package~\cite{mohr2014daubechies}, where we decided to follow the second approach,
% in this way obtaining  support functions which are optimally adapted to their chemical environment and thus yield a very high accuracy. We are in this way able to drastically reduce the size of the subspace in which the density matrix is represented. 

Another important aspect is the choice of the underlying basis which is used to give a numerical representation of the support functions. Ideally, as we would like to efficiently describe localized functions, such a basis set should at the same time feature orthogonality and compact support. This can indeed be offered by Daubechies wavelets~\cite{daubechies1992ten}, making them an ideal basis set for linear scaling calculations. Moreover, wavelets are able to yield a reasonable precision with only moderate values for the numerical grid spacing, thus keeping the number of basis functions relatively small for a systematic approach. In addition the multiresolution properties of wavelets allow for an adaptive resolution and thus to have a finer mesh only in those regions -- close to the atoms -- where it is actually required. 

As for a given system the localized Wannier functions are unknown,  there are two possibilities: either use a larger number of support functions and hope that the most important features of the Wannier functions can be captured in this way, or to use a smaller number of support functions and optimize them in situ (namely within a predefined localization region) to get as close as possible to the Wannier functions. We have recently implemented a linear scaling code based on Daubechies wavelets, distributed in the BigDFT package~\cite{mohr2014daubechies}, where we decided to follow the second approach, in this way obtaining  support functions which are optimally adapted to their chemical environment and thus yield a very high accuracy. Consequently we are in this way able to drastically reduce the size of the subspace in which the density matrix is represented. 
Together with the aforementioned properties of Daubechies wavelets this leads to an optimally reduced number of degrees of freedom and enables us to perform calculations on thousands of atoms while only requiring moderate amounts of memory and CPU time.

%As for a given system the localized Wannier functions are unknown, there are two possibilities: either use a larger number of support functions and hope that the most important features of the Wannier functions can be captured in this way, or to use a smaller number of support functions and optimize them \emph{in situ} (namely within a predefined localization region) to get as close as possible to the Wannier functions. 
%We have recently implemented a linear scaling code based on Daubechies wavelets, distributed in the BigDFT package~\cite{mohr2014daubechies}, where we decided to follow the second approach,
% in this way obtaining  support functions which are optimally adapted to their chemical environment and thus yield a very high accuracy. We are in this way able to drastically reduce the size of the subspace in which the density matrix is represented. 
  
In this paper we will present the progress we have recently made for this linear scaling code.
%based on Daubechies wavelets, distributed in the BigDFT package~\cite{mohr2014daubechies}.
%This approach has the advantage that one does not need to do a tedious fine tuning of various parameters and in particular of the basis, but can straightforwardly get reasonable results using a set of default values.
Thanks to the previously presented ingredients our code offers the ability to perform DFT calculations which at the same time yield an astonishingly high accuracy, are universally applicable, exhibit a linear complexity, and require only a very small amount of computational resources. In particular it offers the advantage that one does not need to do a tedious fine tuning of various parameters and in particular of the basis, but can straightforwardly get reasonable results using a set of default values.

We will show that even with a relatively simple set of input parameters one can calculate energies with an absolute accuracy of the order of \unit[10]{meV/atom} for a large variety of isolated systems. From a computational point of view, as the approach scales linearly with respect to system size, it is possible to use the concept of ``CPU minutes per atom'' to evaluate the computational resources needed for a full DFT calculation.
We will first present the basic principles of our approach and then demonstrate all of the aforementioned properties by various examples.

%Thanks to the aforementioned ingredients we were able to implement a new scheme into the well established BigDFT~\cite{genovese2008daubechies} code which has the ability to perform calculations which at the same yield an astonishingly high accuracy, are universally applicable, exhibit a linear complexity, and require only a very small amount of computational resources. In this paper we will demonstrate all of these properties by first presenting the basic principles of our approach and then illustrating the mentioned features by various examples.

% Even if this Moreover the small number of support functions helps to drastically reduce the size of the subspace in which 

% Moreover the small number of support functions helps to drastically reduce the computational resources which are required, thus allowing to perform calculations of thousands of atoms without requiring huge amounts of memory and CPU time.
% Together with the fact the Daubechies wavelets which are used to give a numerical representation of the support function are an ideal basis for linear scaling calculations, this results in a code which is able to perform linear scaling calculations on large systems with an only moderate demand of resources.

\section{Theory and algorithm}
\subsection{Overview of the algorithm}
Using the formalism based on the density matrix $\mathbf{F}(\mathbf{r},\mathbf{r}')$, the band structure energy, which is the central quantity to be calculated, is given within the framework of Kohn-Sham DFT by the expression
\begin{equation}
 E_{BS} = \left. \int \mathcal{H}(\mathbf{r}')F(\mathbf{r},\mathbf{r}')\right|_{\mathbf{r}=\mathbf{r}'} \mathrm{d}\mathbf{r}'.
\end{equation}
Here $\mathcal{H}(\mathbf{r})$ is the Kohn-Sham Hamiltonian, defined by
\begin{equation}
 \mathcal{H}(\mathbf{r})=-\frac{1}{2}\mathbf{\nabla}^2 + \mathcal{V}_{KS}(\rho(\mathbf{r}),\mathbf{r}) + \mathcal{V}_{PSP}(\mathbf{r}),
\end{equation}
where $\mathcal{V}_{KS}(\rho(\mathbf{r}),\mathbf{r}) = \mathcal{V}_{ext}(\mathbf{r}) + \int \frac{\rho(\mathbf{r}')}{|\mathbf{r}-\mathbf{r}'|}\mathrm{d}\mathbf{r}' + \mathcal{V}_{XC}(\mathbf{r})$ is the Kohn-Sham potential containing the external potential, the Hartree potential and the exchange-correlation potential, and $\mathcal{V}_{PSP}$ is the pseudopotential representing the ions in the system. In BigDFT the pseudopotentials are norm-conserving GTH-HGH~\cite{hartwigsen1998relativistic} pseudopotentials and their Krack variants~\cite{krack2005pseudopotentials}, possibly enhanced by a nonlinear core correction~\cite{willand2013norm-conserving}.

The linear scaling version of BigDFT is based on a separable ansatz for the density matrix. Using so-called support functions $\phi_\alpha(\mathbf{r})$ and a density kernel $\mathbf{K}$ the density matrix can be written as
\begin{equation}
 F(\mathbf{r},\mathbf{r}') = \sum_{\alpha,\beta} \phi_\alpha(\mathbf{r})K_{\alpha\beta}\phi_\beta(\mathbf{r'}).
\end{equation}
The charge density, which enters the Kohn-Sham Hamiltonian, is given by the diagonal part of the density matrix and thus reads
\begin{equation}
 \rho(\mathbf{r}) = \sum_{\alpha,\beta} \phi_\alpha(\mathbf{r})K_{\alpha\beta}\phi_\beta(\mathbf{r}).
\end{equation}
Using the orthonormality relations between the support functions and their duals, $\int\phi_\alpha(\mathbf{r})\tilde{\phi}_\beta(\mathbf{r})\mathrm{d}\mathbf{r} = \delta_{\alpha\beta}$, it follows that the elements of the density kernel are given by
\begin{equation}
 K_{\alpha\beta} = \iint\tilde{\phi}_\alpha(\mathbf{r})F(\mathbf{r},\mathbf{r}')\tilde{\phi}_\beta(\mathbf{r}')\mathrm{d}\mathbf{r}\mathrm{d}\mathbf{r'}.
 \label{eq:kernel_elements_in_terms_of_support_functions}
\end{equation}
Writing analogously the elements of the Hamiltonian matrix as
\begin{equation}
 H_{\alpha\beta} = \int\phi_\alpha(\mathbf{r})\mathcal{H}(\mathbf{r})\phi_\beta(\mathbf{r}')\mathrm{d}\mathbf{r},
\end{equation}
it follows that the band structure energy is given by
\begin{equation}
 E_{BS} = \tr(\mathbf{K}\mathbf{H}).
 \label{eq:band_structure_energy_as_trace}
\end{equation}
The algorithm thus consists of two parts: finding a good set of support functions $\phi_\alpha$, which give rise to the Hamiltonian matrix $\mathbf{H}$, and determining the density kernel $\mathbf{K}$ in the dual (and never explicitly calculated) basis $\tilde{\phi}_\alpha(\mathbf{r})$. In order to reach linear scaling, the support functions must be localized and the density kernel sparse. This can be achieved by introducing cutoff radii around the center of each support function and setting to zero all components which lie outside of the region; for the density kernel an element $K_{\alpha\beta}$ is set to zero if the centers of $\phi_\alpha$ and $\phi_\beta$ are farther away than the sum of the kernel cutoffs of the regions $\alpha$ and $\beta$ -- these kernel cutoffs actually correspond to the cutoffs of the dual support functions (cf. Eq.~\ref{eq:kernel_elements_in_terms_of_support_functions}) and are in general a bit larger than the support function cutoffs themselves, see the discussion in Sec.~\ref{sec:Optimization of the support function and the density kernel}. The support functions are again expanded in terms of an underlying wavelet basis, as is explained in more details in Ref.~\onlinecite{mohr2014daubechies}.

\begin{figure}
  \includegraphics[width=0.3\textwidth]{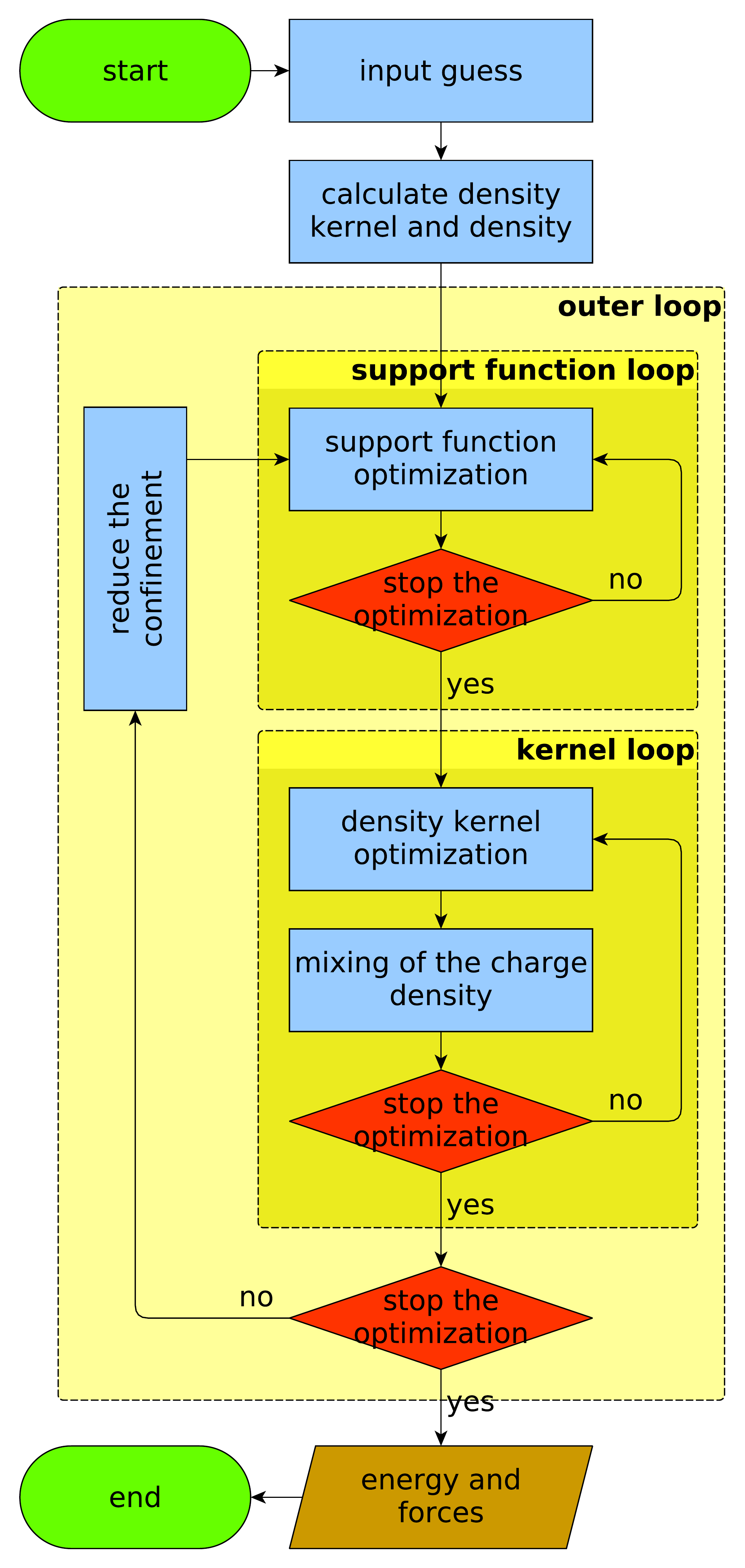}
  \caption{Flowchart of the linear scaling algorithm. It consists of a double loop structure: in the inner first loop the support functions are optimized with a fixed density kernel, in the second inner loop the density kernel is optimized using a fixed set of support functions. These two loops are then iterated in an outer loop until overall convergence is reached.}
  \label{fig:flowchart}
\end{figure}
A flowchart of the algorithm is shown in Fig.~\ref{fig:flowchart}. It consists of a nested double loop structure, with one outer loop and two inner loops. In the first inner loop, the support functions are optimized using a fixed kernel and thus also using a fixed potential and Hamiltonian. Therefore it is not advisable to perform too many iterations in this loop, but rather to exit even if the convergence threshold has not been reached yet. In the second inner loop, the density kernel is optimized using a fixed set of support functions. In contrast to the first inner loop, this second one is done self-consistently, i.e.\ the Hamiltonian is updated in each iteration. These two loops are then iterated in the outer loop until overall convergence is reached.

\subsection{Optimization of the support functions and density kernel}
\label{sec:Optimization of the support function and the density kernel}
The support functions are optimized by minimizing a target function. Ideally this target function should lead to strongly localized support functions which at the same time yield a very high accuracy. For the latter property the correct quantity to be minimized is the band structure energy of Eq.~\eqref{eq:band_structure_energy_as_trace}. For the first property the choice is not unique; one possibility would be
\begin{equation}
 \Omega = \sum_\alpha \braket{\phi_\alpha|\mathcal{H}_\alpha|\phi_\alpha},
\end{equation}
where $\mathcal{H}_\alpha$ is the Kohn-Sham Hamiltonian including a confinement, i.e. $\mathcal{H}_\alpha = \mathcal{H} + c_\alpha (\mathbf{r}-\mathbf{R}_\alpha)^4$, with $\mathbf{R}_\alpha$ being the center of the localization region.
In order to combine the two properties we define our target function as
\begin{equation}
 \Omega = \sum_\alpha K_{\alpha\alpha} \braket{\phi_\alpha|\mathcal{H}_\alpha|\phi_\alpha} + \sum_{\beta\ne\alpha}K_{\alpha\beta}\braket{\phi_\alpha|\mathcal{H}|\phi_\beta},
\end{equation}
where the prefactor for the confinement is smoothly reduced during the run, as explained in more detail in Ref.~\onlinecite{mohr2014daubechies}. In this way we have a strong confinement in the beginning, leading to a decent localization, and still obtain a high precision as we correctly minimize the band structure energy in the end. Since the decrease of the confinement is done automatically -- taking into account the properties of the system --, this procedure is universally applicable without the need for any fine tuning.

In contrast to other linear scaling codes employing the same support function approach~\cite{skylaris2005introducing,bowler2010calculations,vandevondele2005quickstep,soler2002the} we decided to impose an additional constraint and to keep the support functions approximately orthogonal.
In other terms, we optimize $\phi_\alpha$ such that the overlap matrix
\begin{equation}
\mathbf{S}_{\alpha\beta} = \langle \phi_\alpha | \phi_\beta \rangle
\end{equation}
is close to the identity matrix.
As the kernel cutoff is related to the extension of the dual support functions (see Eq.\eqref{eq:kernel_elements_in_terms_of_support_functions}),
%the inverse overlap matrix has to be sparse if we want to preserve the sparsity of $\mathbf{K}$.
the sparsity of $\mathbf{K}$ is governed by the cutoff implicitly used for these dual functions.
%This has several advantages.
%Firstly our support functions form a systematic set in this way. Secondly this will allow us to choose smaller values for the kernel cutoff compared to non-orthogonal support functions -- this is due to the fact that the kernel cutoff is related to the cutoff of the dual support functions (cf. Eq.~\eqref{eq:kernel_elements_in_terms_of_support_functions}) whose extent gets larger the more localized the non-dual support functions are; b
%Keeping the support functions orthogonal leads to a degree of sparsity which is comparable to that of the overlap matrix itself. The dual and non-dual entities are thus identical and the cutoffs for the support functions and
%the density kernel matrix elements can be chosen along the same lines.
% The same argument also applies to the inverse of the overlap matrix.
By choosing the support functions to be orthogonal the dual and non-dual entities are identical and the cutoffs for the support functions and
the density kernel matrix elements can consequently be chosen along the same lines, leading to a degree of sparsity for the kernel which is comparable to that of the overlap matrix itself.

In addition, quasi-orthogonal support functions lead to an overlap matrix which is close to the identity and whose inverse can thus be cheaply calculated using polynomial expansions. 
%Moreover the same arguments as for the density kernel are also valid for the inverse of the overlap matrix, i.e.\ 
%The use of orthogonal functions leads to a degree of sparsity which is comparable to that of the overlap matrix itself. 
This property is in particular important for the Fermi Operator Expansion~\cite{goedecker1994efficient,goedecker1995tight-binding} (FOE) procedure that we use to determine the density kernel. The efficiency of this approach relies heavily on the sparsity of the matrix $\mathbf{S}^{-1/2}\mathbf{H}\mathbf{S}^{-1/2}$. Even if $\mathbf{H}$ is very sparse thanks to the cutoff radii of the support functions, this property is much less pronounced for $\mathbf{S}^{-1/2}$ in the case of non-orthogonal support functions. Thus, if one wants to keep a high degree of sparsity also for the matrix product, it is indispensable to work with a set of orthogonal support functions. 

In spite of all these benefits, it must be stated that orthogonality and locality are in general two contradicting properties; consequently a strict enforcement of the orthogonality might lead to convergence problems. Therefore we only perform an explicit orthogonalization in the very beginning; 
in the following the orthogonality is only approximately conserved by the use of a Lagrange multiplier in the support function gradient. This is enough to keep the overlap matrix diagonally dominant -- the off-diagonal elements being typically of the order of $0.1$ -- and thus to maintain the aforementioned benefits. 
		
For the determination of the density kernel the code offers several possibilities, each one having its particular strengths and thus areas of application. In this paper we always used the FOE method, as it is the only method that allows to perform calculations for very large systems. For more details we again refer to Ref.~\onlinecite{mohr2014daubechies}.

\section{Assessment of the accuracy}
\label{sec:Assessment of the accuracy}
In order to assess the precision of our code, it has to be compared with a reference. We use as such the cubic version of the code~\cite{genovese2008daubechies}, which has been shown to yield very accurate results~\cite{willand2013norm-conserving}. Thus, by demonstrating that the linear version of BigDFT is able to reproduce the results of the cubic scaling version, we can be sure to have reached a very high level of precision. All calculations were done using free boundary conditions and the PBE~\cite{perdew1996generalized} functional unless otherwise stated.
\subsection{Comparison of energies and forces}
\label{sec:Comparison of energies and forces}
\begin{figure}
 \includegraphics[width=0.19\textwidth]{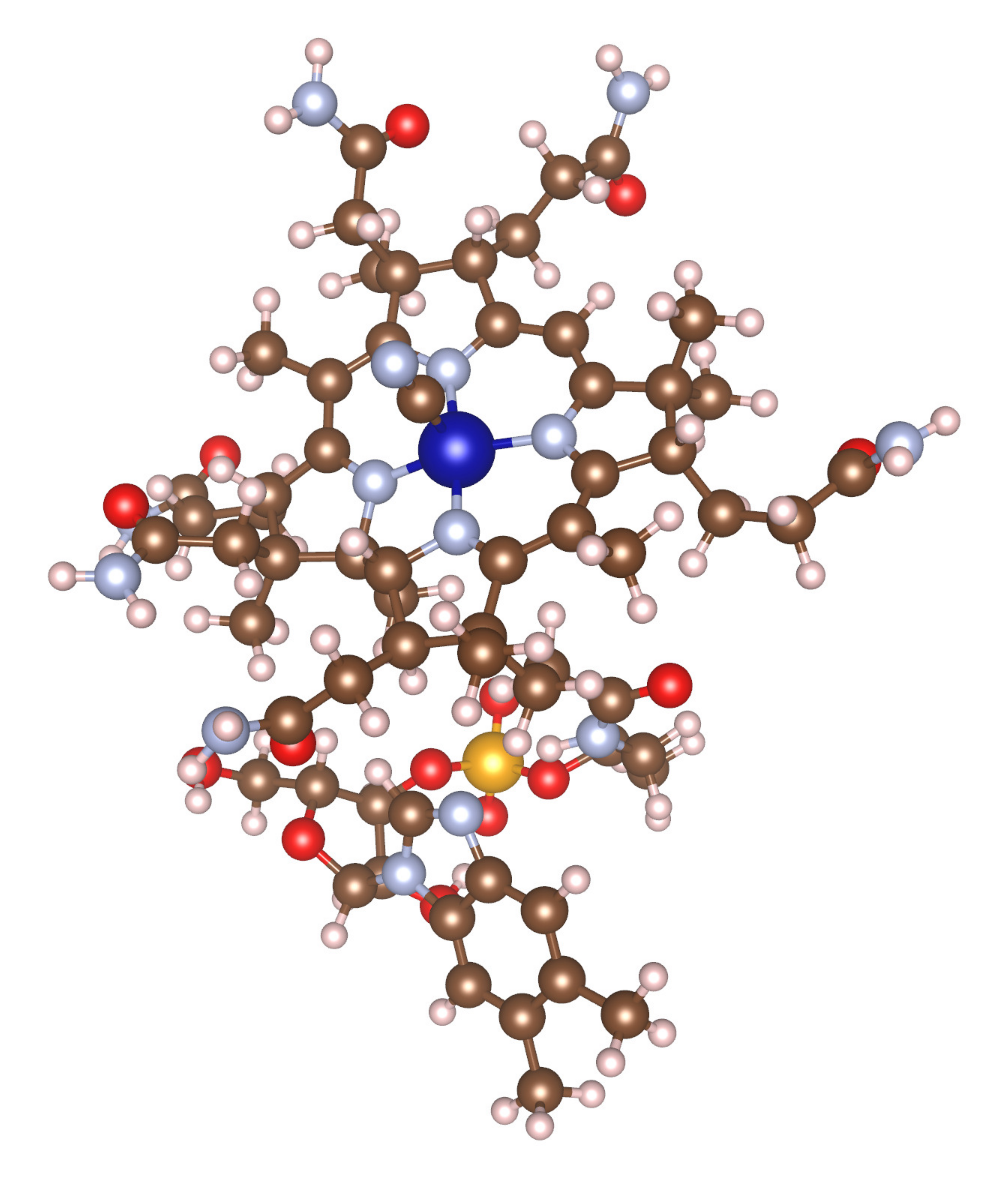}
 \hspace{-20pt}
 \includegraphics[width=0.19\textwidth]{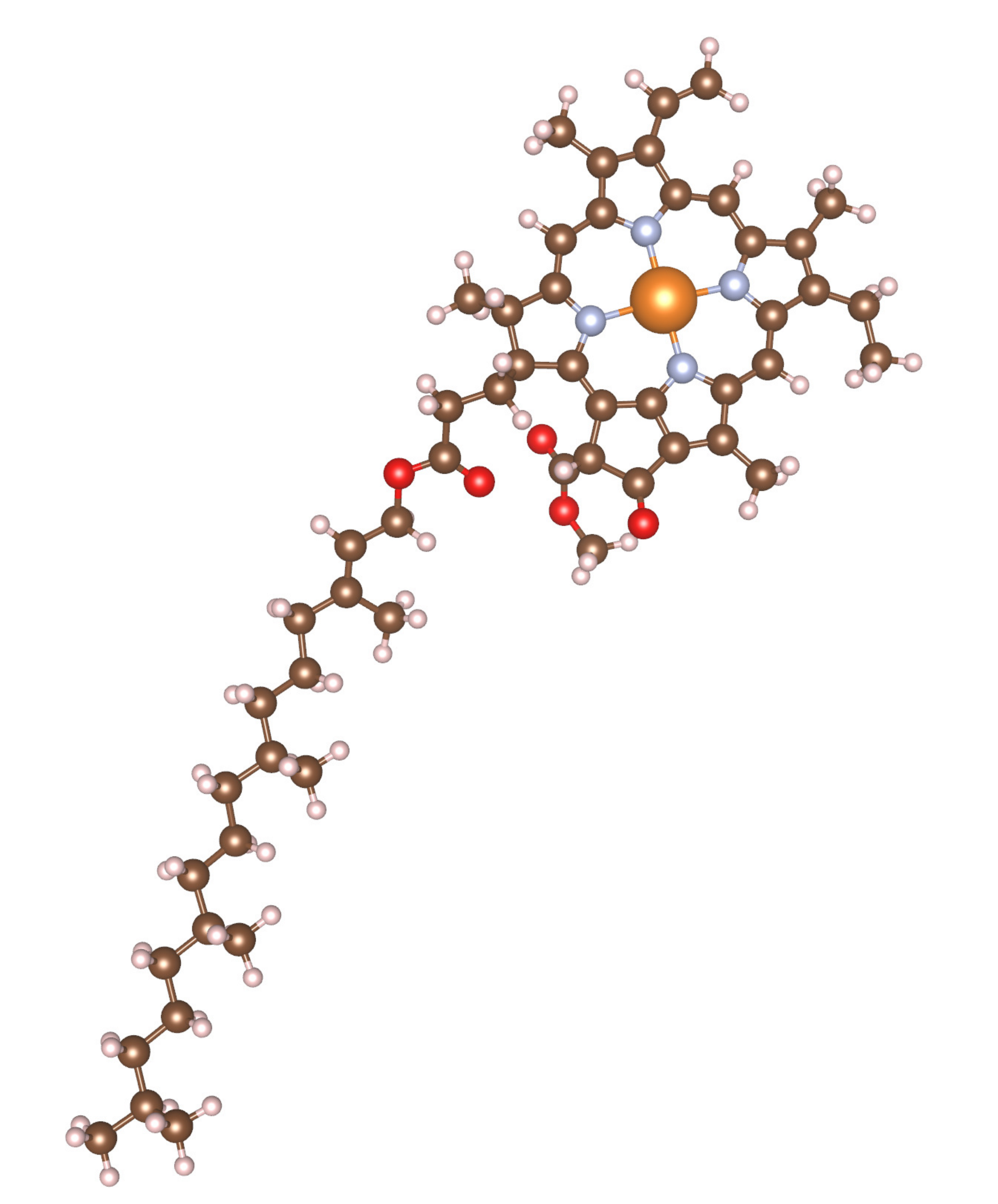}
 \hspace{-30pt}
 \includegraphics[width=0.13\textwidth]{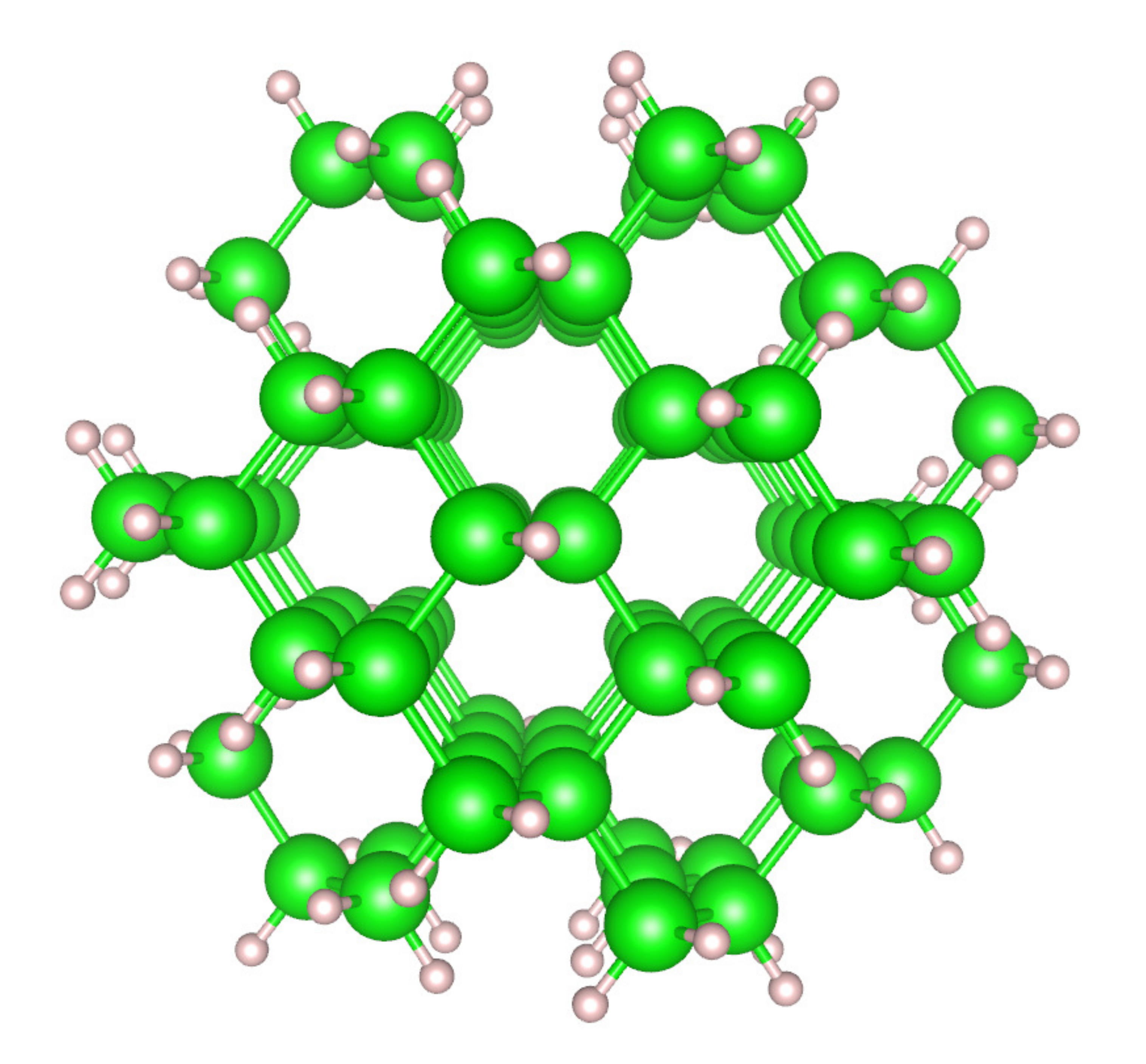} \\
%  \hspace{5pt}
 \includegraphics[width=0.09\textwidth]{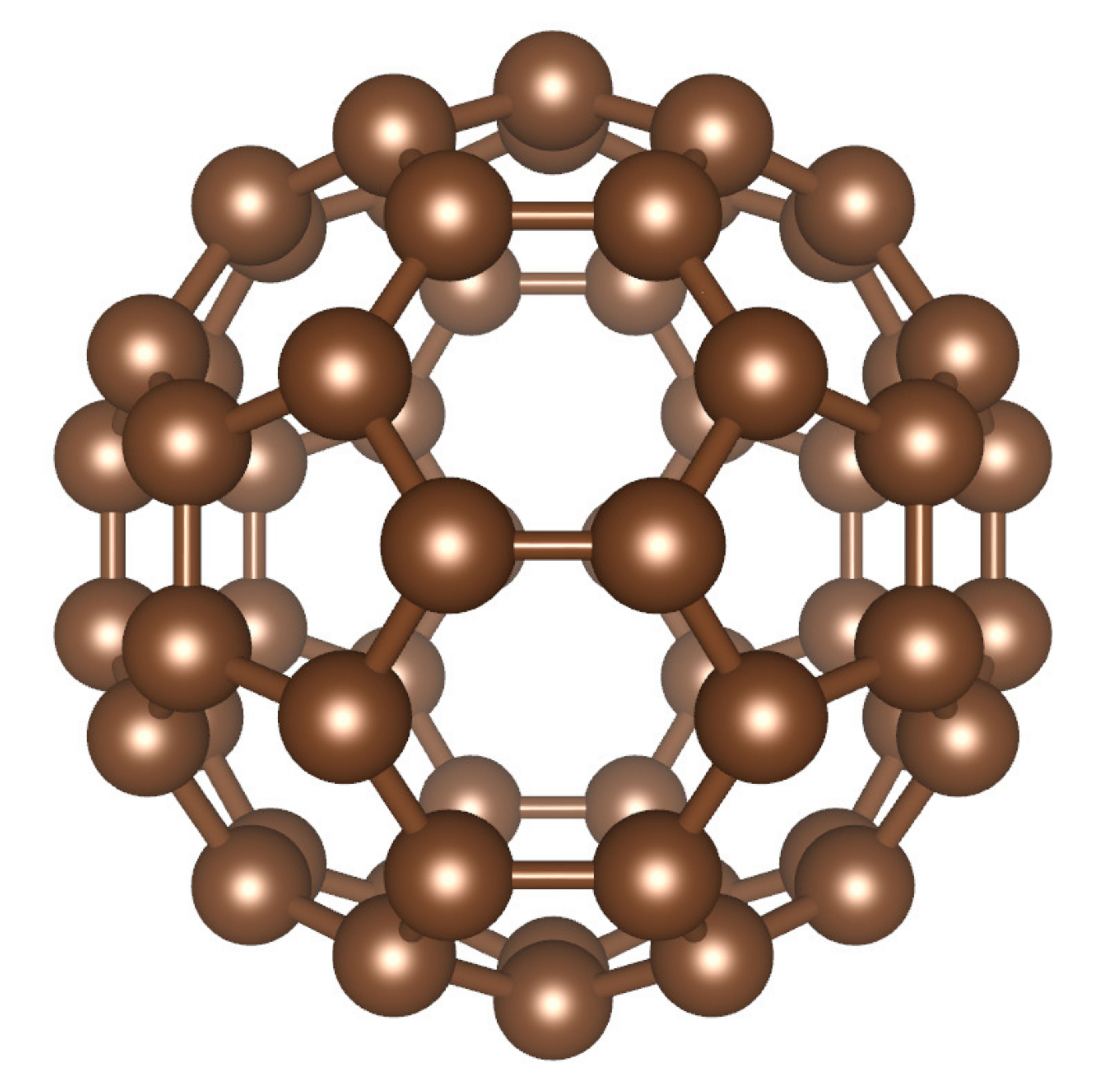}
 \hspace{5pt}
 \includegraphics[width=0.16\textwidth]{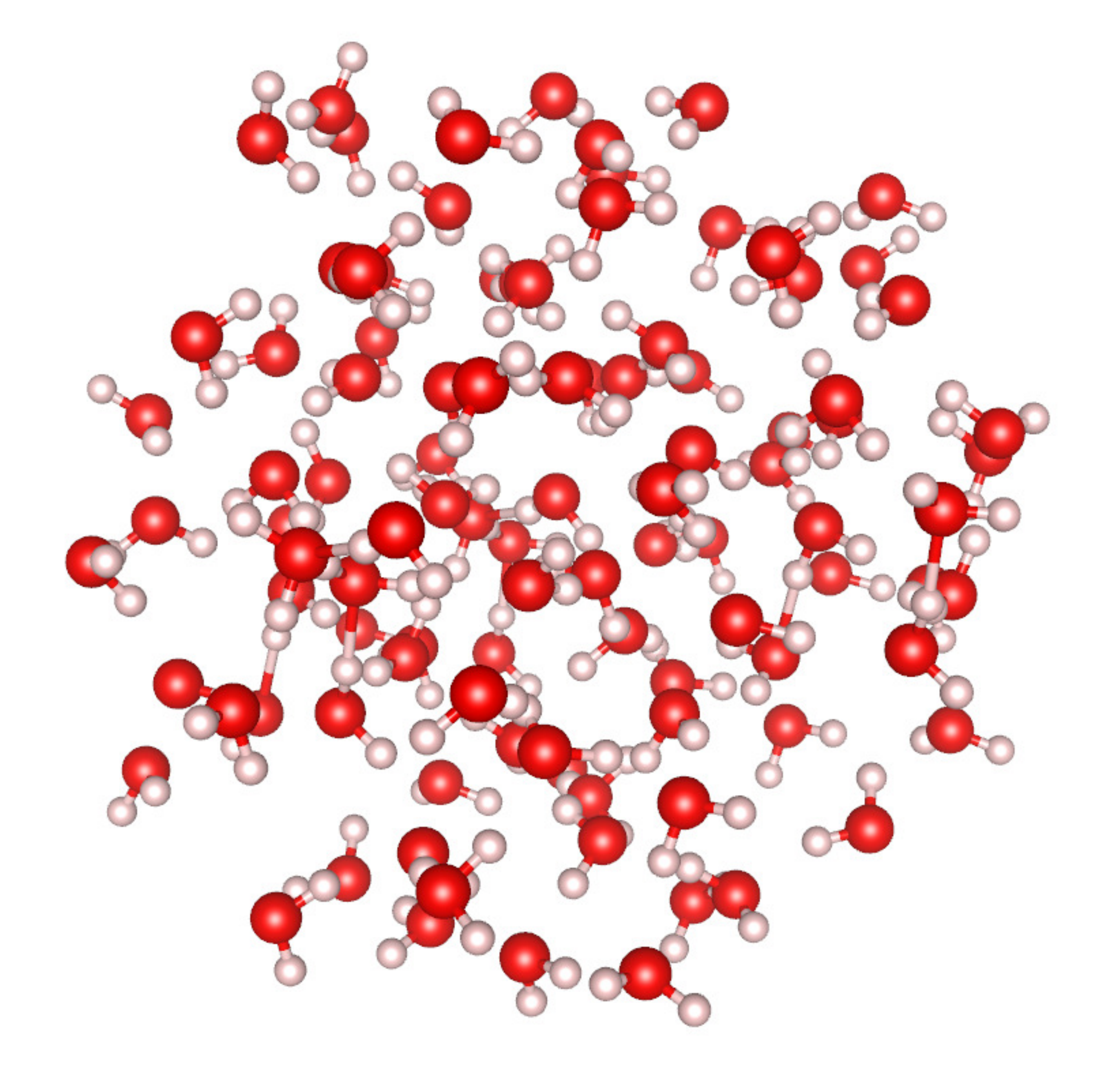}
 \hspace{5pt}
 \includegraphics[width=0.19\textwidth]{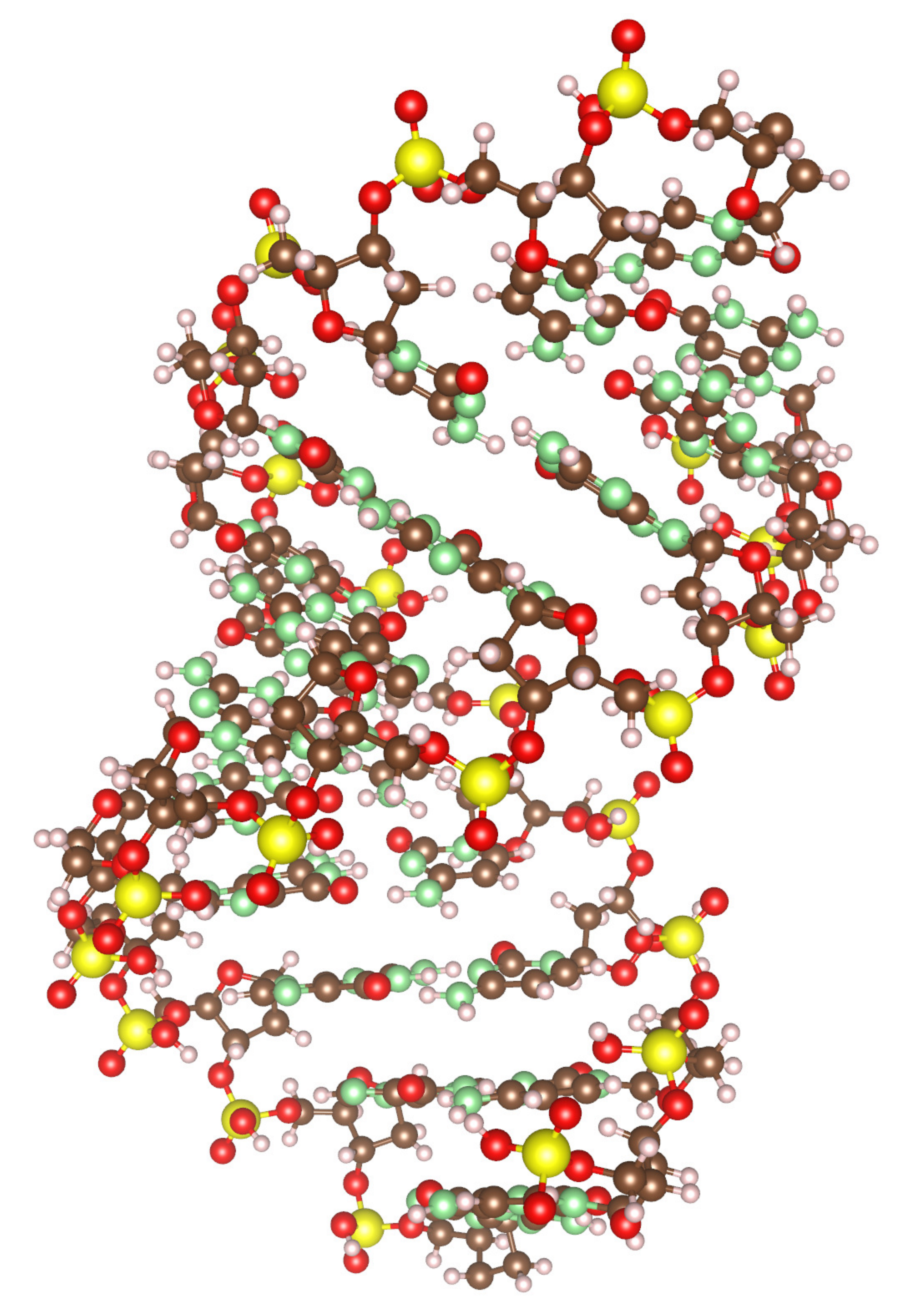}
 \caption{Illustration of the six systems used for the comparison of the linear and cubic energies and forces: Vitamin \ce{B12}, Chlorophyll, \ce{Si87H76}, \ce{C60}, \ce{(H2O)100}, DNA.}
 \label{fig:systems_used_for_comparison}
\end{figure}
One of the most important features of our code is the ability to yield a high level of precision without the need for doing a tedious fine tuning of the parameters. As an illustration of this property, we show the energies and forces calculated with both the linear and cubic scaling version for six systems with rather different electronic structures; they are depicted in Fig.~\ref{fig:systems_used_for_comparison}. The chosen parameters were very similar for all systems, with the main difference being the various cutoff radii used for the support function regions of the different atom kinds. To keep it simple, the choice was just based on the row in which the element appears in the periodic table: \unit[5.5]{bohr} for elements of the first row, \unit[6.3]{bohr} for those of the second row, and \unit[7.0]{bohr} for those of the third row. The kernel cutoff was set to \unit[9.0]{bohr}; an automatic adjustment for technical reasons was done by the code if the value was smaller than the 
cutoff radius of the support function plus 8 times the grid spacing, see Ref.~\onlinecite{mohr2014daubechies} for details. The number of support functions per atom 
was 1 for \ce{H}, 9 for \ce{Si}, and 4 for all the other elements. These parameters seem to be fairly universal, meaning that any other system can be readily calculated without the need for any adjustments.

\begin{table*}
%  \scriptsize
 \centering
 \begin{tabular}{l r r r r r r r r}
  \hline\hline
                       & \multicolumn{2}{c}{cubic} & & \multicolumn{2}{c}{linear}  & & \multicolumn{2}{c}{difference} \\
                          \cline{2-3} \cline{5-6} \cline{8-9}
                       & \multicolumn{1}{c}{energy}        & \multicolumn{1}{c}{force norm}         & & \multicolumn{1}{c}{energy}        & \multicolumn{1}{c}{force norm}         & &
                         \multicolumn{1}{c}{energy}        & \multicolumn{1}{c}{force norm} \\
                       & \multicolumn{1}{c}{\scriptsize hartree} & \multicolumn{1}{c}{\scriptsize hartree/bohr} & & \multicolumn{1}{c}{\scriptsize hartree} & \multicolumn{1}{c}{\scriptsize hartree/bohr} & &
                         \multicolumn{1}{c}{\scriptsize meV/atom}& \multicolumn{1}{c}{\scriptsize hartree/bohr} \\
     \hline
      Vitamin \ce{B12} & $-926.78$                        & $2.13\cdot 10^{-3}$                   & & $-926.71$                        & $1.93 \cdot 10^{-2}$                  & &
                         $11.43$                          & $1.72\cdot 10^{-2}$ \\
      Chlorophyll      & $-476.70$                        & $3.05\cdot 10^{-3}$                   & & $-476.64$                        & $1.24 \cdot 10^{-2}$                  & &
                         $12.33$                          & $9.38\cdot 10^{-3}$ \\
      \ce{Si87H76}     & $-386.79$                        & $7.80\cdot 10^{-4}$                   & & $-386.69$                        & $1.01 \cdot 10^{-2}$                  & &
                         $17.20$                          & $9.27\cdot 10^{-3}$ \\
      \ce{C60}         & $-341.06$                        & $2.69\cdot 10^{-4}$                   & & $-341.02$                        & $7.36 \cdot 10^{-3}$                  & &
                         $17.23$                          & $7.09\cdot 10^{-3}$ \\
      \ce{(H2O)100}    & $-1722.99$                       & $5.23\cdot 10^{-1}$                   & & $-1722.87$                       & $5.26 \cdot 10^{-1}$                  & &
                         $10.89$                          & $2.80\cdot 10^{-3}$\\
      DNA              & $-4483.12$                       & $5.60\cdot 10^{-1}$                   & & $-4482.84$                       & $5.63 \cdot 10^{-1}$                  & &
                         $10.69$                          & $2.40\cdot 10^{-3}$\\
   \hline\hline
 \end{tabular}
 \caption{Comparison of the energies and forces for different systems, calculated using the linear and cubic versions. Linear energies have an offset of about \unit[10]{meV/atom} with respect to the cubic version, and the deviations of the linear forces from their cubic counterparts is of the order of \unit[$10^{-3}$]{hartree/bohr}. The first four systems were close to their geometrical ground state, whereas the last two were unrelaxed. All systems were calculated with similar parameters.}
 \label{tab:comparison_linear_cubic}
\end{table*}
The results are summarized in Tab.~\ref{tab:comparison_linear_cubic}. As can be seen, it is possible to get with these input parameters a fairly constant energy offset between the linear and cubic versions of about \unit[10]{meV/atom}. The difference of the force norm is typically of the order of \unit[$10^{-3}$]{hartree/bohr}. These are more than reasonable values in view of the fact that they were obtained without any fine tuning. Last but not least, it is worth noting that support functions are optimized such that Pulay forces~\cite{pulay1969ab} are absent in our approach~\cite{mohr2014daubechies}, and the evaluation of the forces is therefore straightforward.

\subsection{Calculations of energy differences}
Absolute energies are always somewhat arbitrary as they are only defined up to an additive constant. Therefore energy differences are more meaningful as this ambiguity vanishes. For the linear scaling version there is in addition the benefit that the offset caused by the finite cutoff radii cancels to a large degree and it is thus possible to get very close to the exact result. The value of \unit[10]{meV/atom} mentioned in the previous section is the error in the absolute energy and can thus be considered as an upper bound; in fact it is however possible to obtain a much higher accuracy than one might think at first sight.
%Moreover, it must be stressed that the value of \unit[10]{meV/atom} mentioned in the previous section is the error in the absolute energy and thus an upper bound. Typically one is interested in energy differences rather than absolute energies; in this case the energy offset cancels to a large degree and the obtained accuracy is much higher than one might think from the presented values. For the linear scaling version there is in addition the benefit that the offset caused by the finite cutoff radii cancels to a large degree and it is thus possible to get very close to the exact result.

\begin{figure}
 \includegraphics[width=0.2\textwidth]{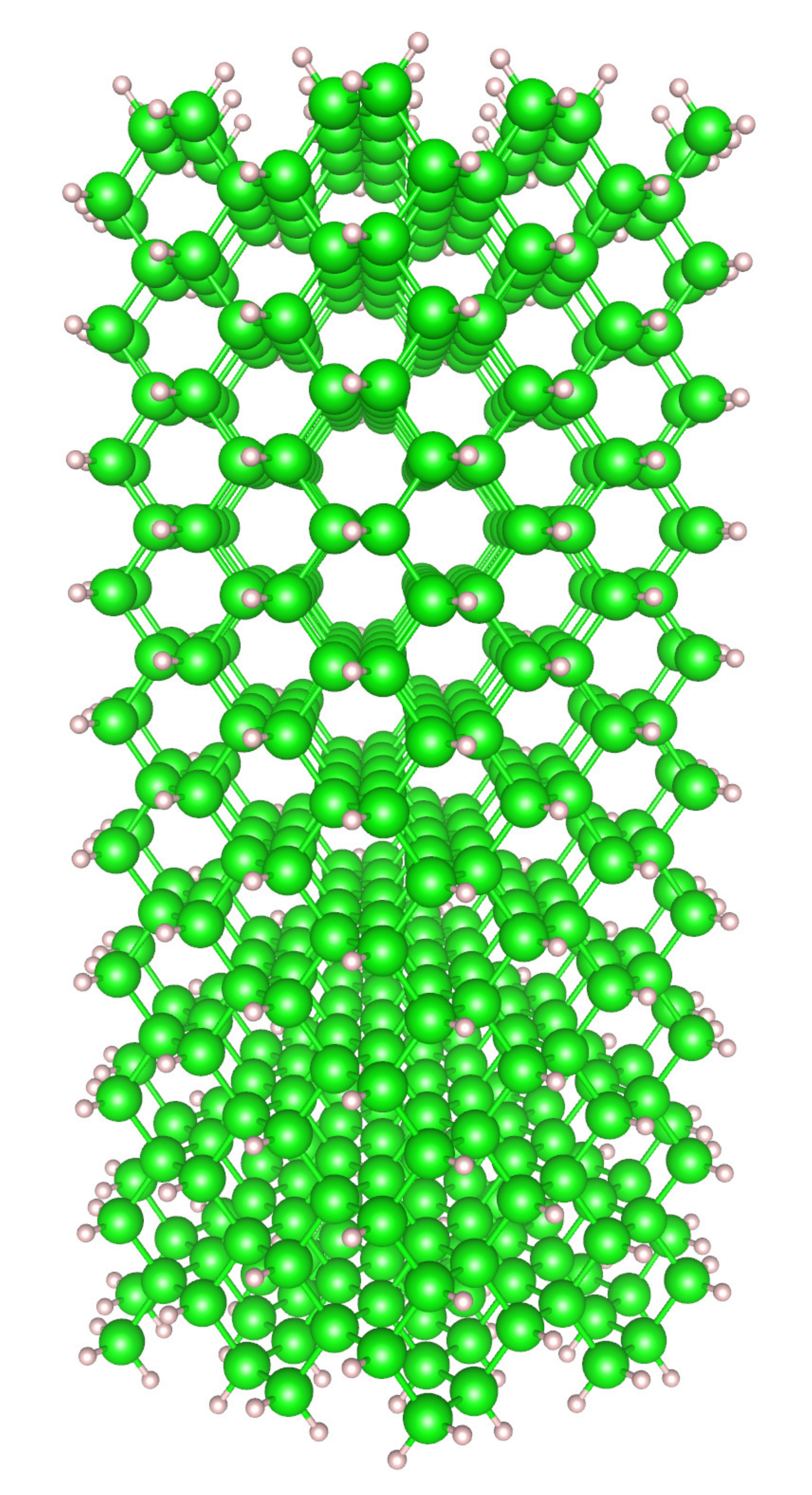}
 \hspace{6pt}
  \includegraphics[width=0.2\textwidth]{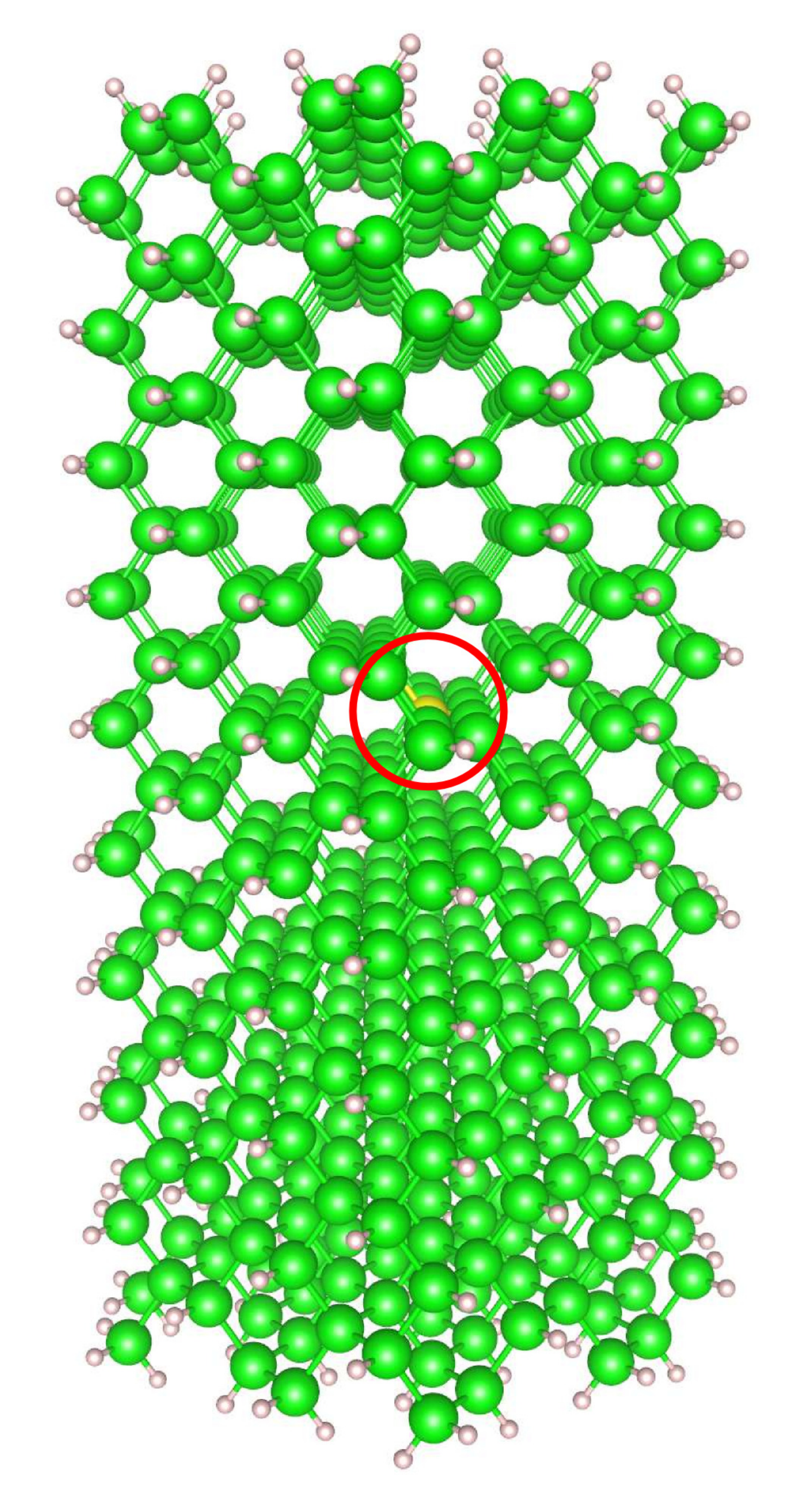}
  \caption{The two silicon wires (consisting of 660 atoms) which are used to benchmark the calculation of energy differences. Whereas the one on the left side is a pure wire, there is a defect in the one on the right side with one silicon atom having been replaced by a phosphorus atom. The red circle highlights this substitutional atom.}
  \label{fig:silicon_wires}
\end{figure}

As an example, we calculated the energy difference between a hydrogen-passivated silicon wire in its pure state and another one which contains a defect. As a defect we chose a substitutional atom; one of the silicon atoms was replaced by phosphorus. These systems have been used as a case study to determine the binding energy of impurities in semiconductors using charged DFT calculations, see Ref.~\onlinecite{PhysRevB.81.161301}. The two configurations are depicted in Fig.~\ref{fig:silicon_wires}. Apart from the fact that silicon is a rather delicate system, this defect is even more challenging as it requires the comparison of two systems with a different number of electrons. The calculations were performed with the analogous parameters as those used for the benchmarks in Sec.~\ref{sec:Comparison of energies and forces}, again demonstrating their universal character. From the results in Tab.~\ref{tab:energy_differences_Si_wire} it becomes clear that energy differences can indeed be calculated with a very high precision. Whereas the offset between the linear and cubic version in the absolute energy is about \unit[11.2]{eV} -- which is about \unit[17]{meV/atom}, in agreement with the results of Sec.~\ref{sec:Comparison of energies and forces} --, the discrepancy between the two version becomes as little as \unit[0.11]{eV} for the energy difference between the pure and the impure wire. Comparing this mismatch with the correct value of \unit[71.8]{eV} one gets a relative error of only 0.15\%.

\begin{table}
%  \scriptsize
 \centering
 \begin{tabular}{l r r r}
 \hline\hline
%      & \multicolumn{1}{c}{cubic} & & \multicolumn{2}{c}{linear} \\
%      \cline{2-2} \cline{4-5}
                             & \multicolumn{1}{c}{pure}     & \multicolumn{1}{c}{impurity}   & \multicolumn{1}{c}{difference} \\
                             & \multicolumn{1}{c}{\scriptsize eV} & \multicolumn{1}{c}{\scriptsize eV}   & \multicolumn{1}{c}{\scriptsize eV} \\
   \hline
    cubic               & -53179.7148                  & -53251.5162                    & 71.8014 \\
    linear              & -53168.4248                  & -53240.3371                    & 71.9123 \\
    difference          &     11.2900                  &     11.1791                    &  0.1109 \\
    relative difference &           -                  &           -                    &  0.15\% \\
  \hline\hline
 \end{tabular}
 \caption{Energy differences between the pure and the impure Si wire, for both the linear and the cubic version. The energy offset between the linear and the cubic version cancels to a large degree and thus yields a very accurate result for the energy difference.}
 \label{tab:energy_differences_Si_wire}
\end{table}

\subsection{Geometry optimizations}
\begin{figure}
 \includegraphics[width=0.48\textwidth]{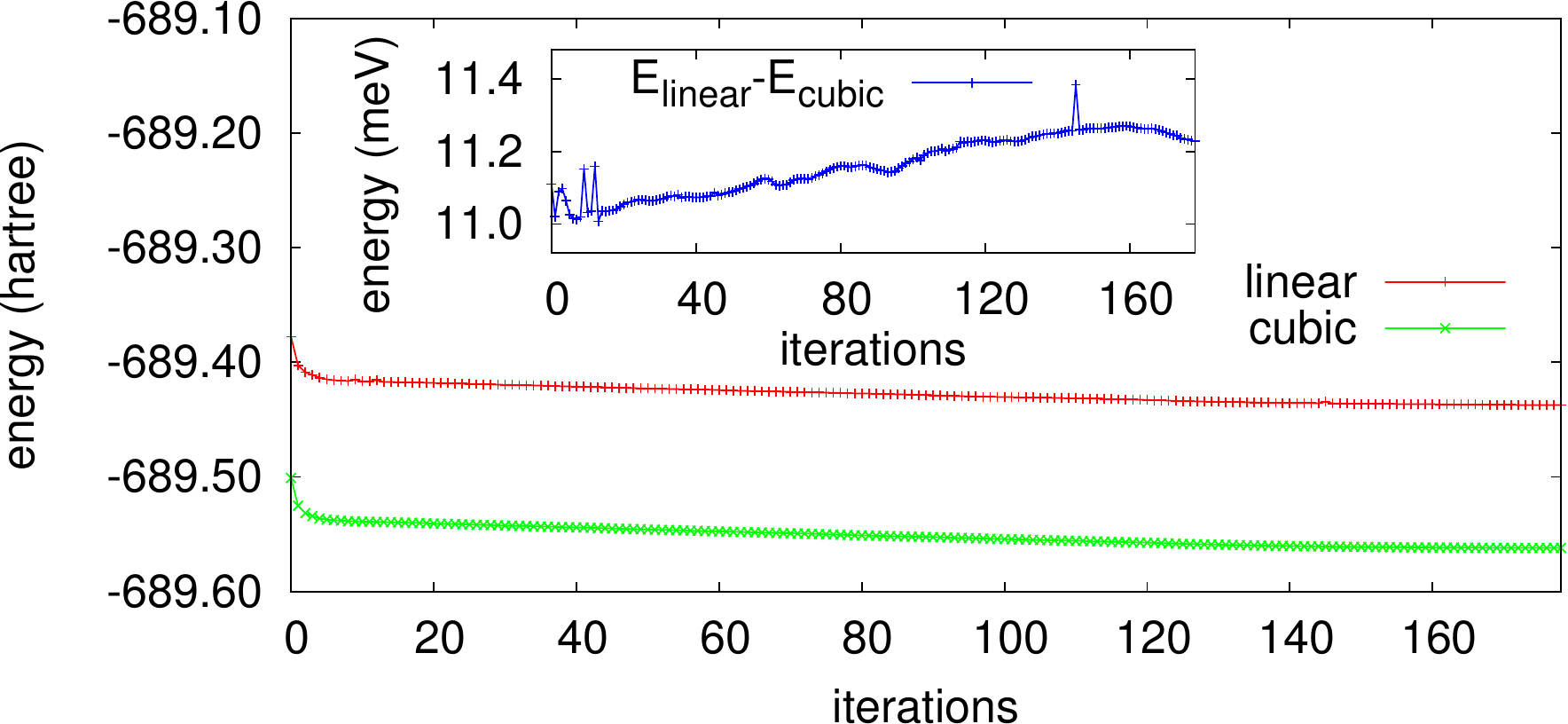}\\
  \vspace{6pt}
 \includegraphics[width=0.48\textwidth]{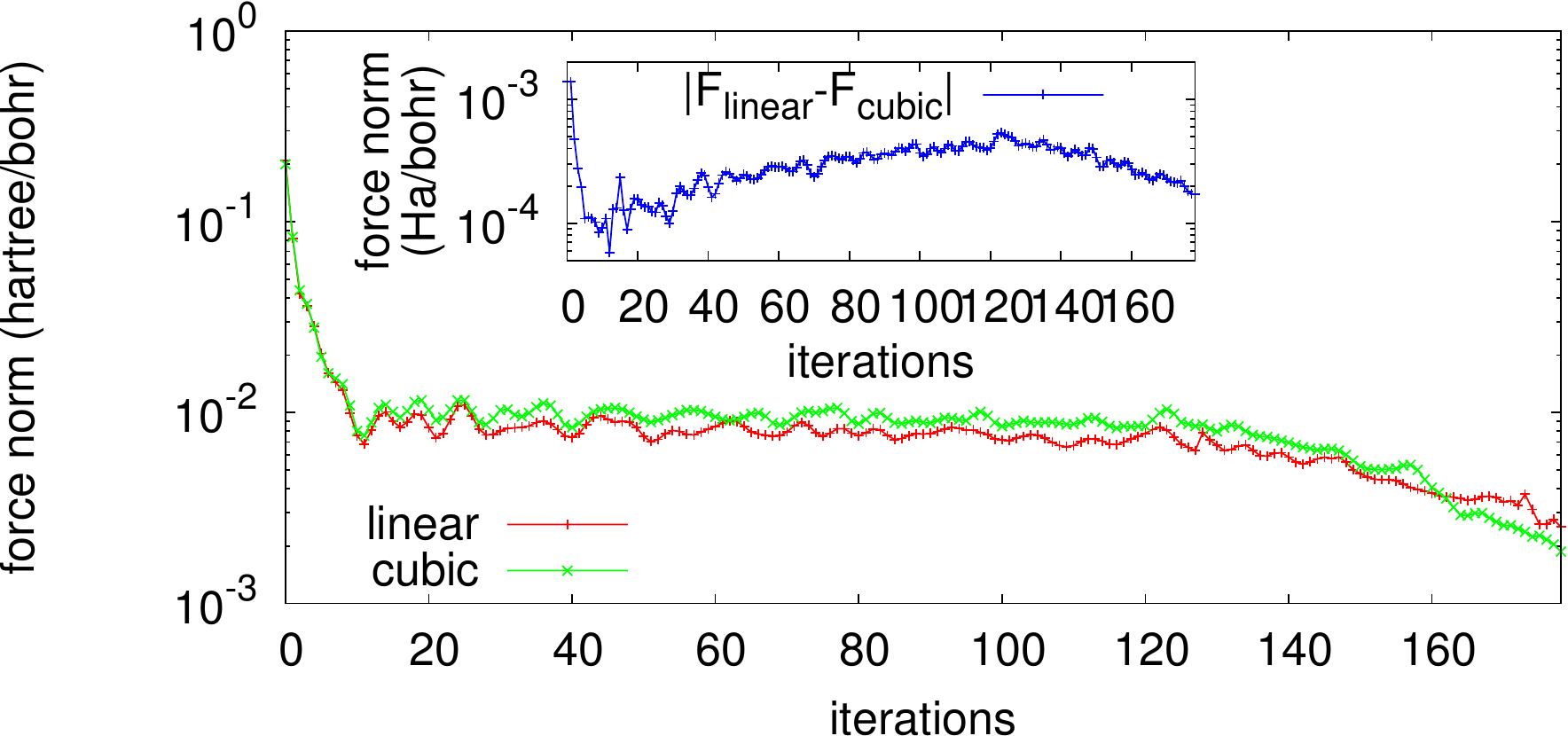}
 \caption{Evaluation of the energy (upper panel) and the forces (lower panel) as a function of the number of iterations in the geometry optimization. For the energy, the offset remains rather constant throughout the entire run, leading to two parallel curves; as can be seen from the inset, the offset only varies within less than \unit[1]{mev/atom}. The curves for the forces are more or less superposed, with a difference of the order of \unit[$10^{-4}$]{hartree/bohr}. The large plateau in the middle where the force remains constant is due to the fact that an adjustment of the bond lengths can only start at the two endpoints of the alkane and then slowly propagate like a wave towards the center of the molecule.}
 \label{fig:geometry_optimization}
\end{figure}

A good test to check at the same time the accuracy of the energies and the forces is to perform a short geometry optimization for both the linear and the cubic version, starting from the same non-equilibrated structure. If the forces calculated by the linear version are accurate enough, this should lead to identical trajectories and thus to a parallel evolution of the energies and the forces. As an example we took an alkane consisting of 302 atoms. The results are shown in Fig.~\ref{fig:geometry_optimization}. As can be seen, the offset in the energy between the linear and the cubic version remains more or less constant throughout the entire geometry optimization, and is again of the order of \unit[10]{meV/atom}. 
%Obviously the difference slightly increases as the run progresses, due to the fact that two trajectories with slightly different initial conditions (which is the case here due to the slight noise in the forces introduced by the finite cutoff radii) deviate exponentially. However the variation is only of the order of ten meV and thus under control. 
In addition, the forces are pretty much identical, with variations of the order of only \unit[$10^{-4}$]{hartree/bohr}. Together this shows that the two trajectories performed by the linear and the cubic versions are identical, demonstrating the high precision of the forces calculated by the linear version. Again it must be stressed that these results were obtained with the same standard set of parameters which had been used in the previous examples.

\subsection{Energetic ordering}
The statement that our approach is universally applicable does not mean that it can blindly be applied to any system. There are situations where the ``noise'' introduced by the finite cutoff radii is larger than the ``signal'' one is looking for; in this case the results of our approach -- as of any method relying on such a truncation -- must be interpreted with care. By universally applicable we rather mean that we can straightforwardly -- i.e.\ without the need of doing a tedious fine tuning of the input parameters and the basis set -- apply our method to those cases where the ``signal to noise ratio'' is large enough even when finite cutoff radii are used. 
%This is analogous to the case of DFT by itself; there are situations where this approach is not able to yield the correct result, and thus every method based on this theory will fail.

\begin{figure}
 \includegraphics[width=0.07\textwidth]{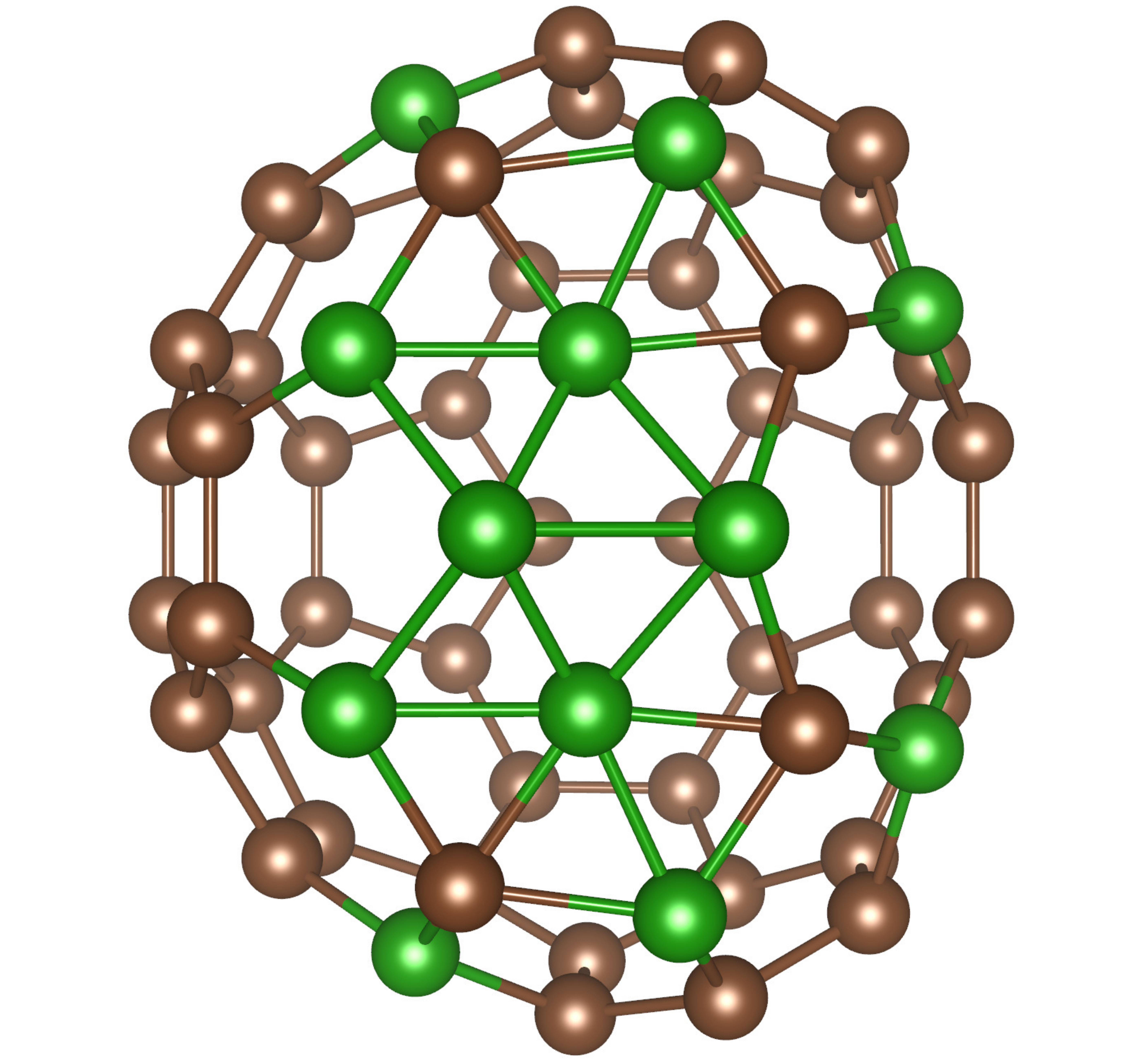}
 \includegraphics[width=0.07\textwidth]{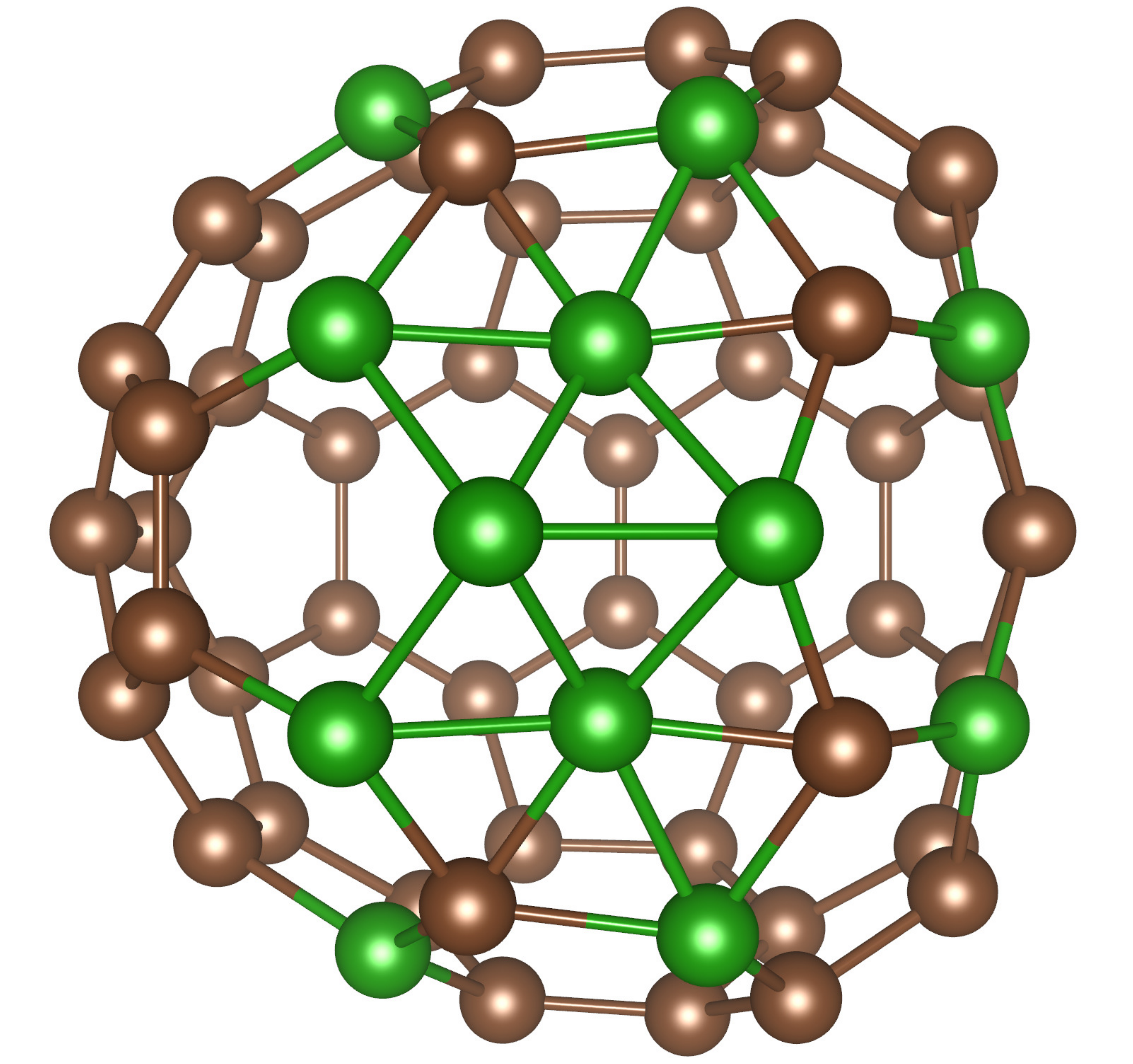}
 \includegraphics[width=0.07\textwidth]{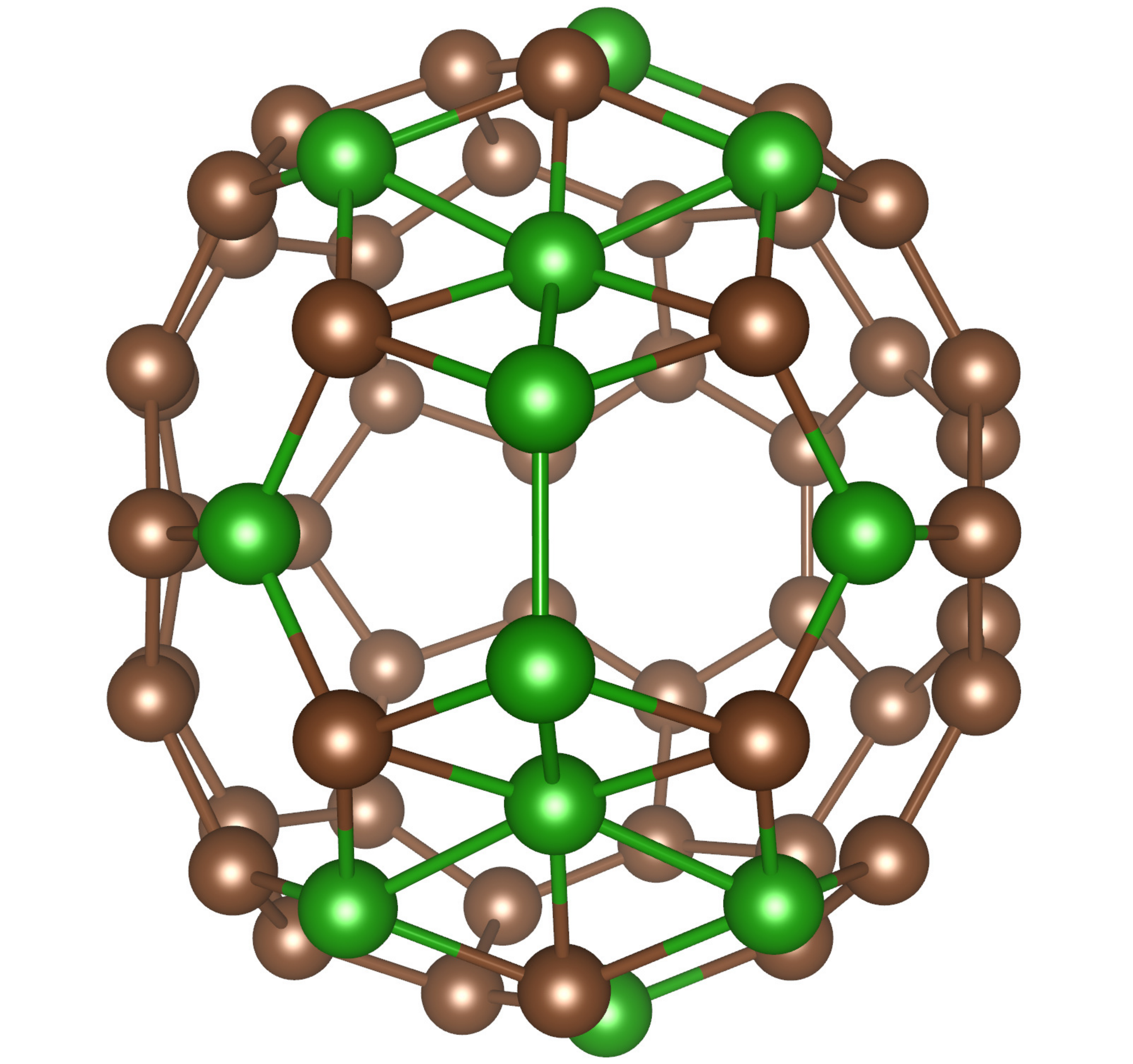}
 \includegraphics[width=0.07\textwidth]{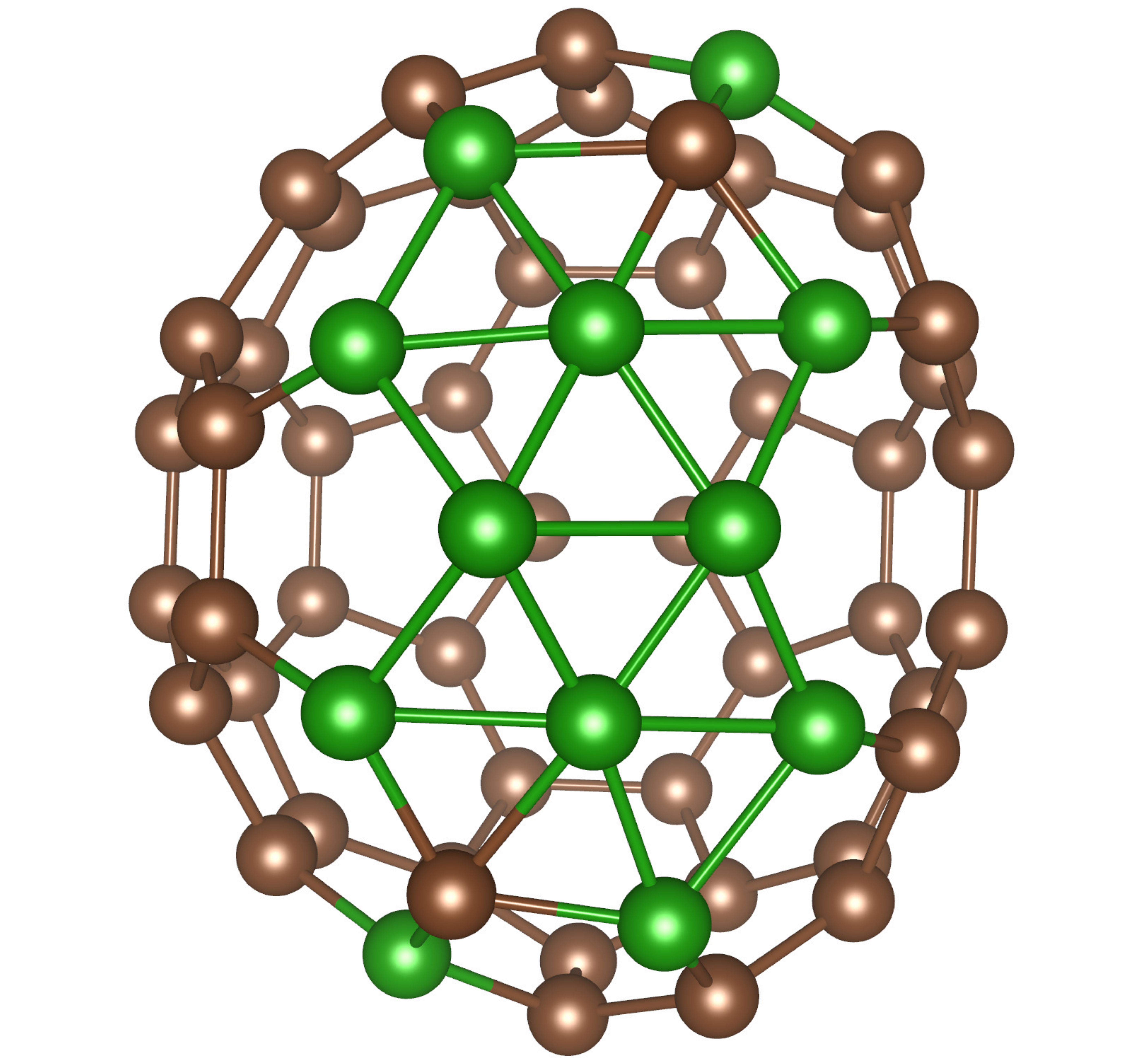}
 \includegraphics[width=0.07\textwidth]{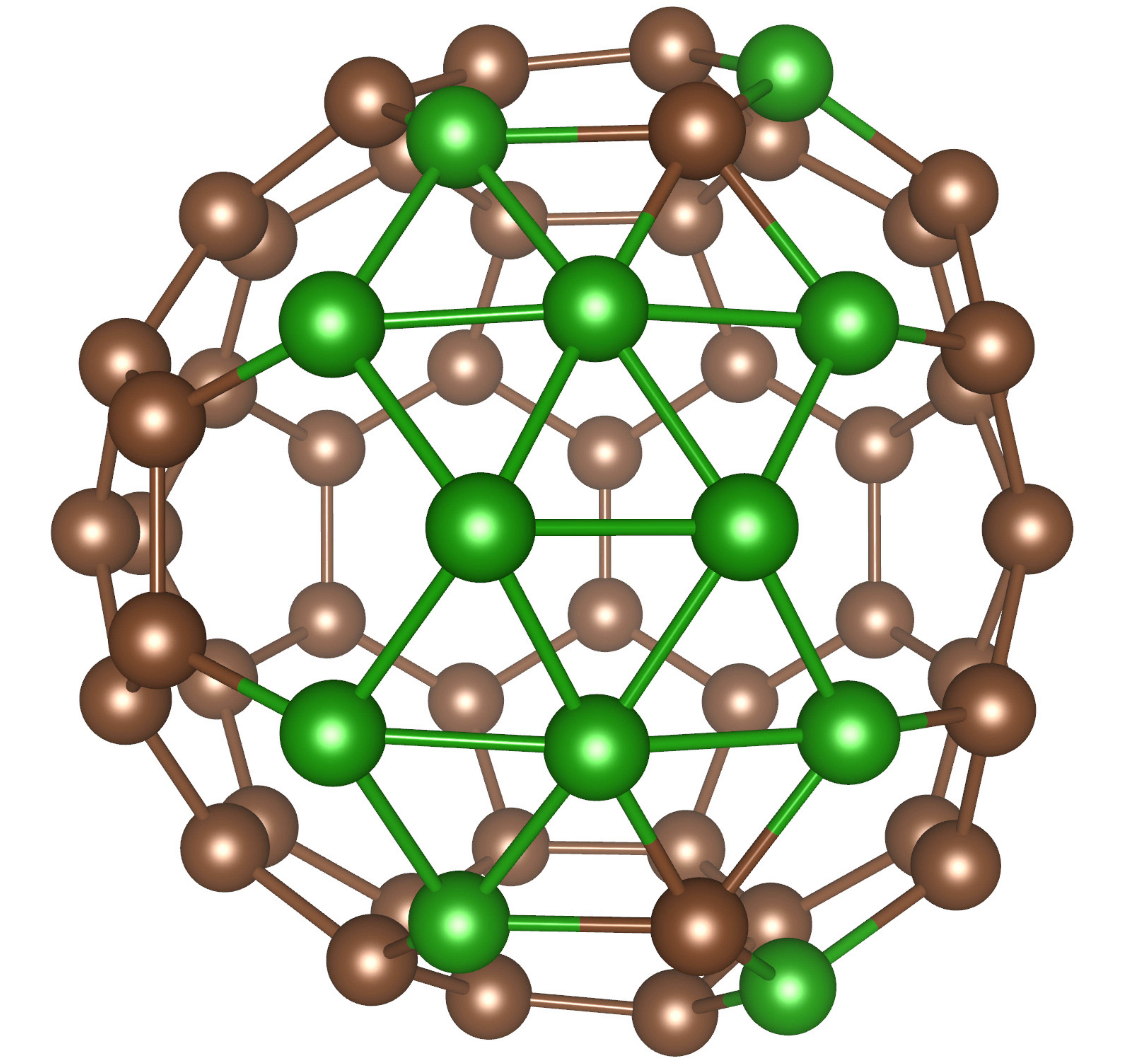}
 \includegraphics[width=0.07\textwidth]{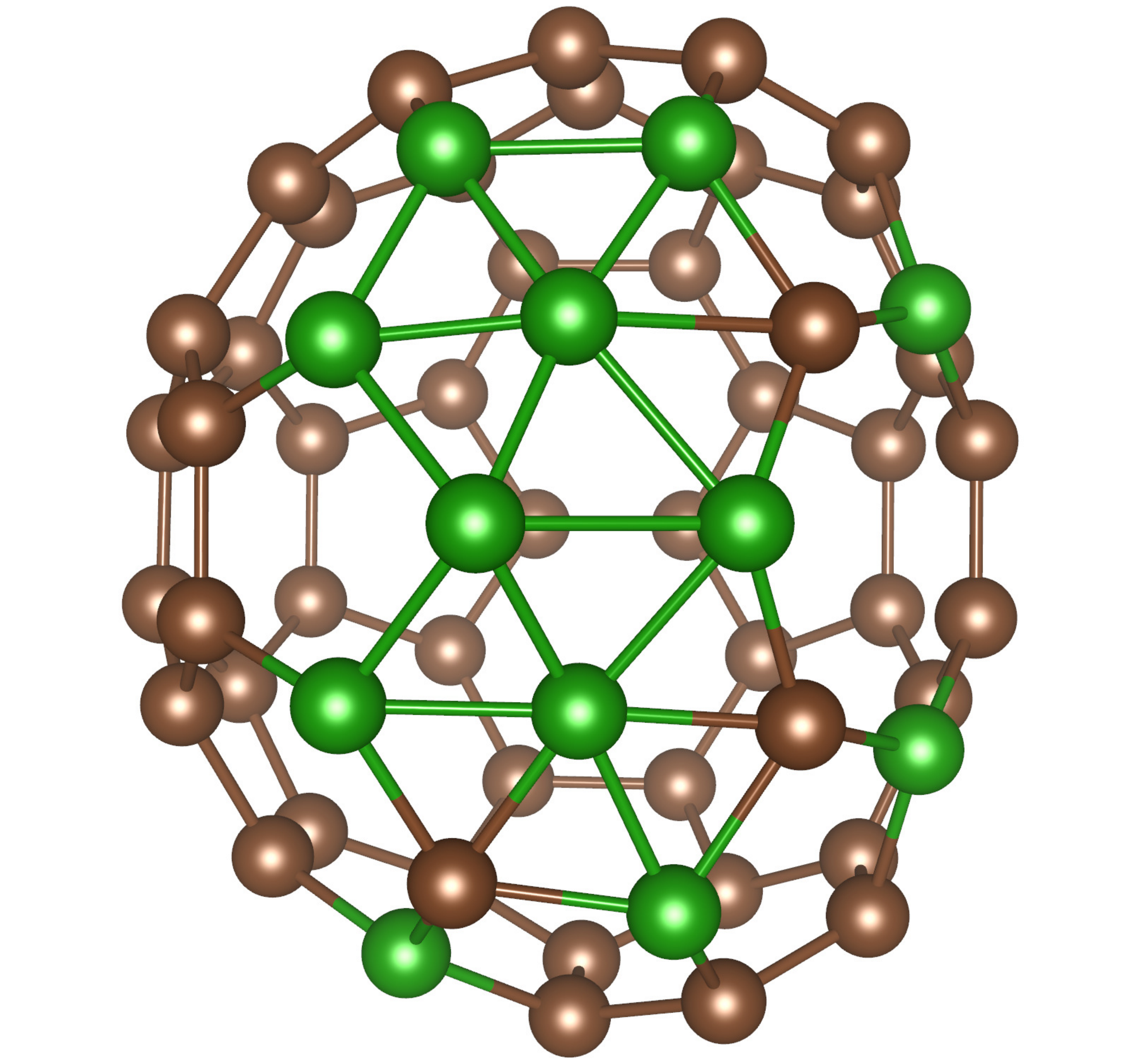}\\
 \vspace{6pt}
 \includegraphics[width=0.07\textwidth]{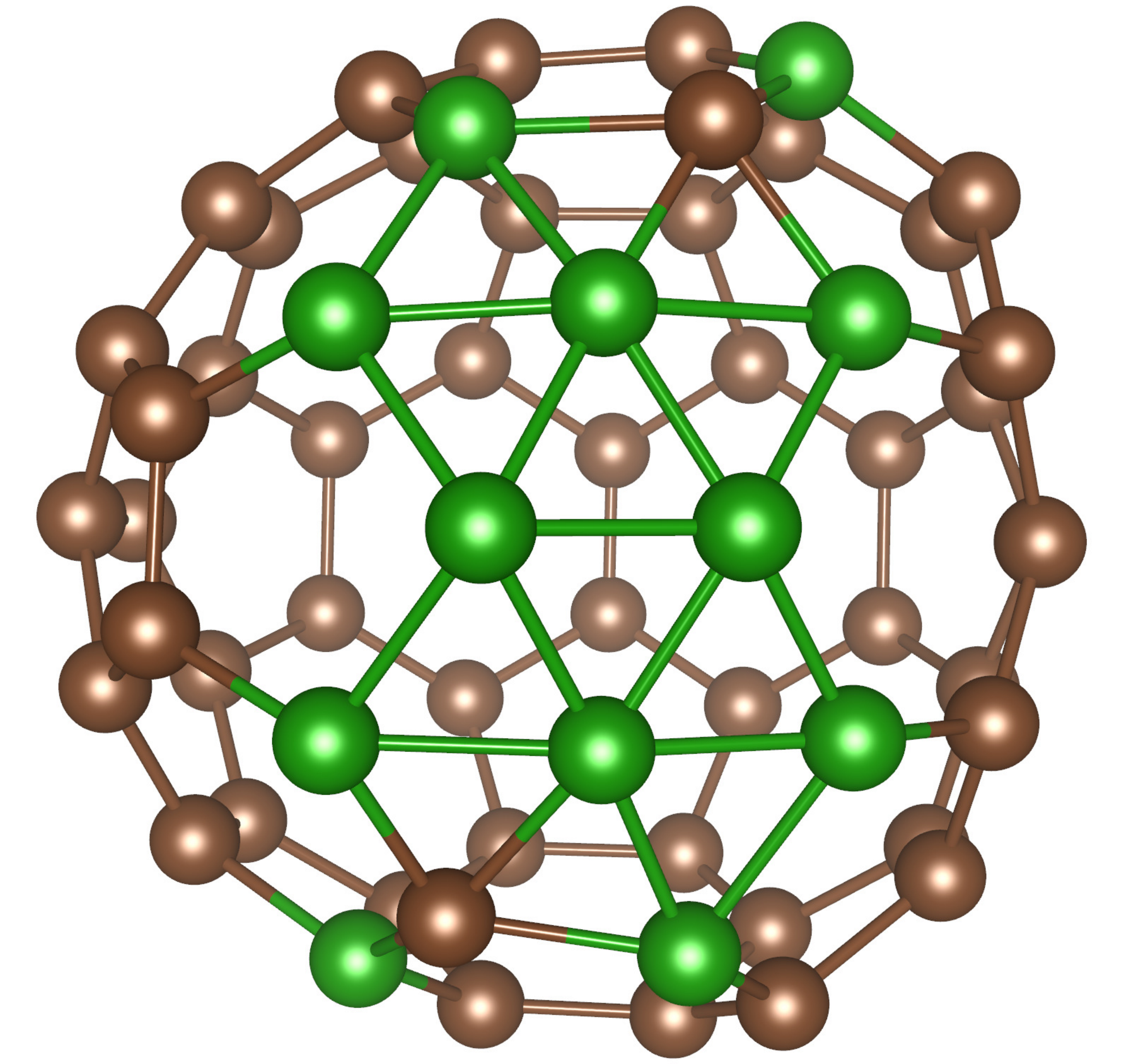}
 \includegraphics[width=0.07\textwidth]{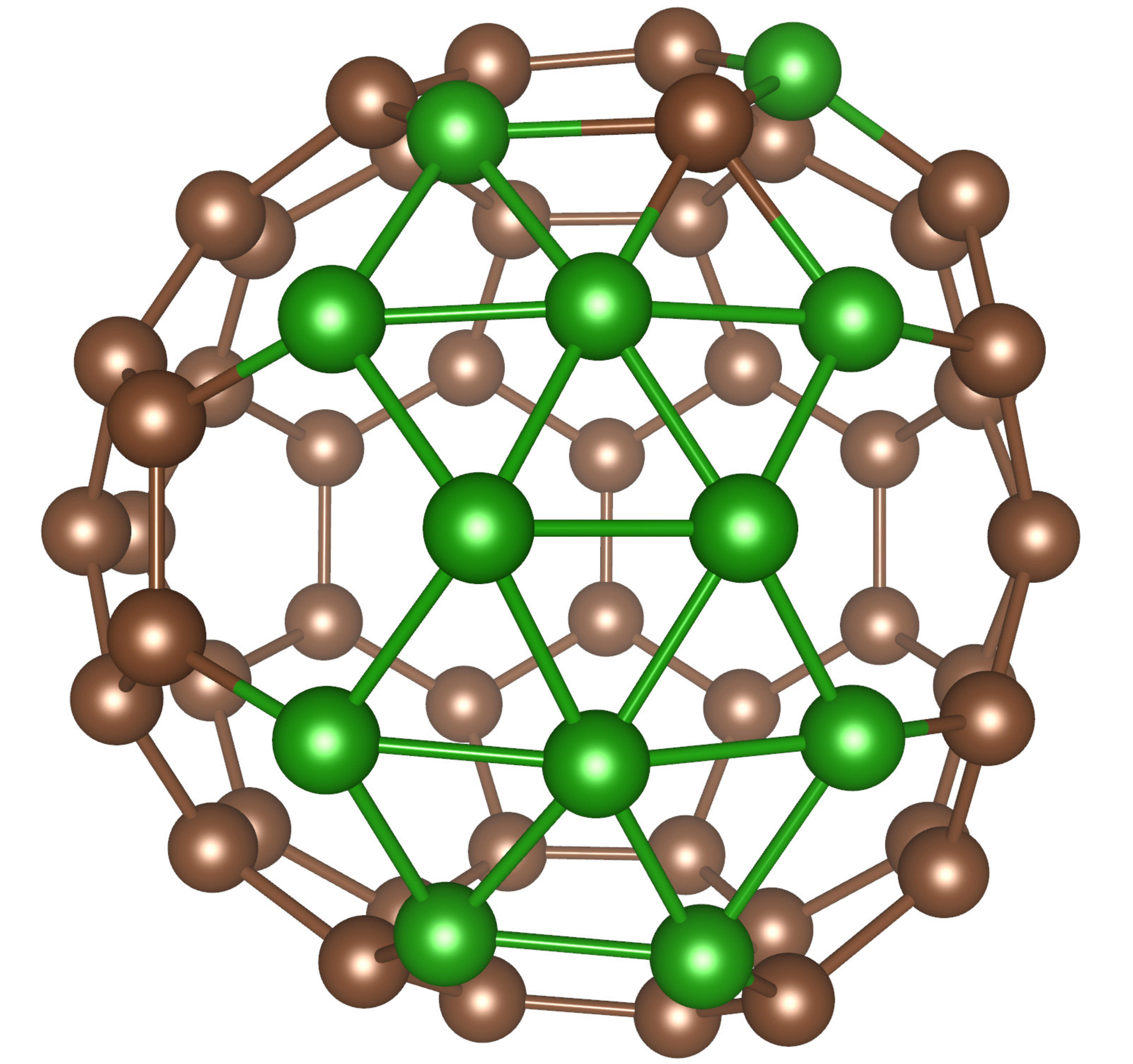}
 \includegraphics[width=0.07\textwidth]{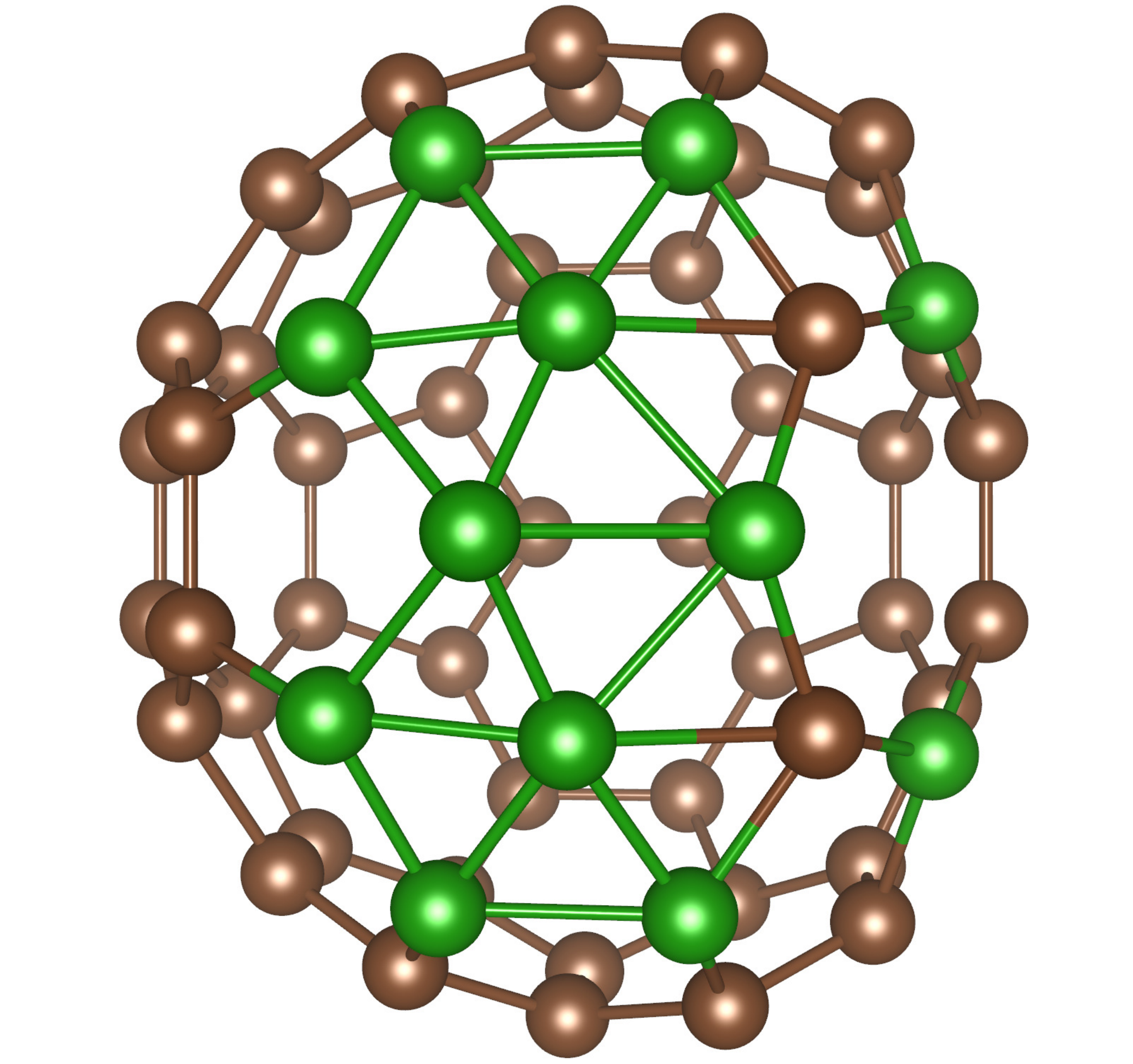}
 \includegraphics[width=0.07\textwidth]{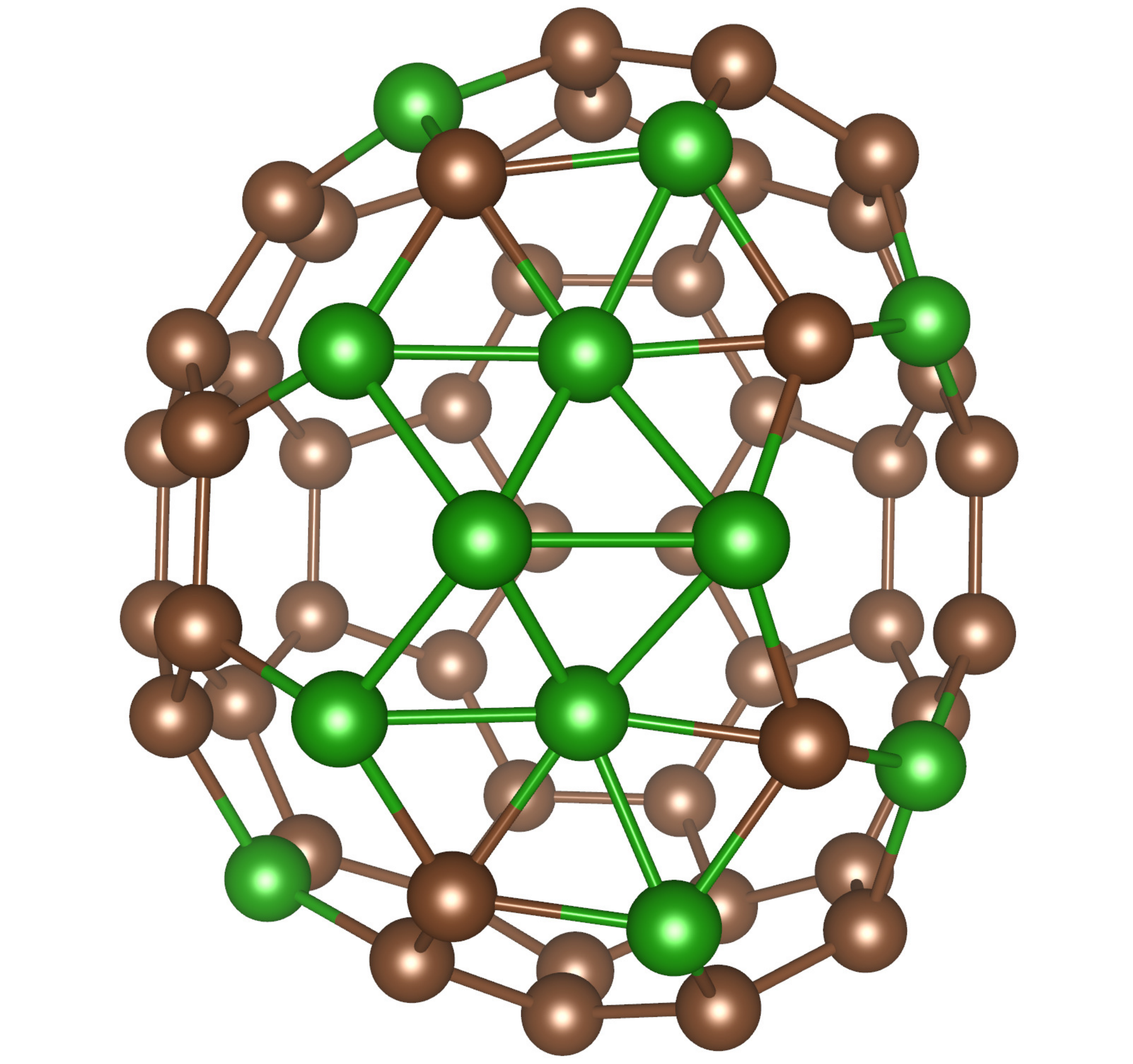}
 \includegraphics[width=0.07\textwidth]{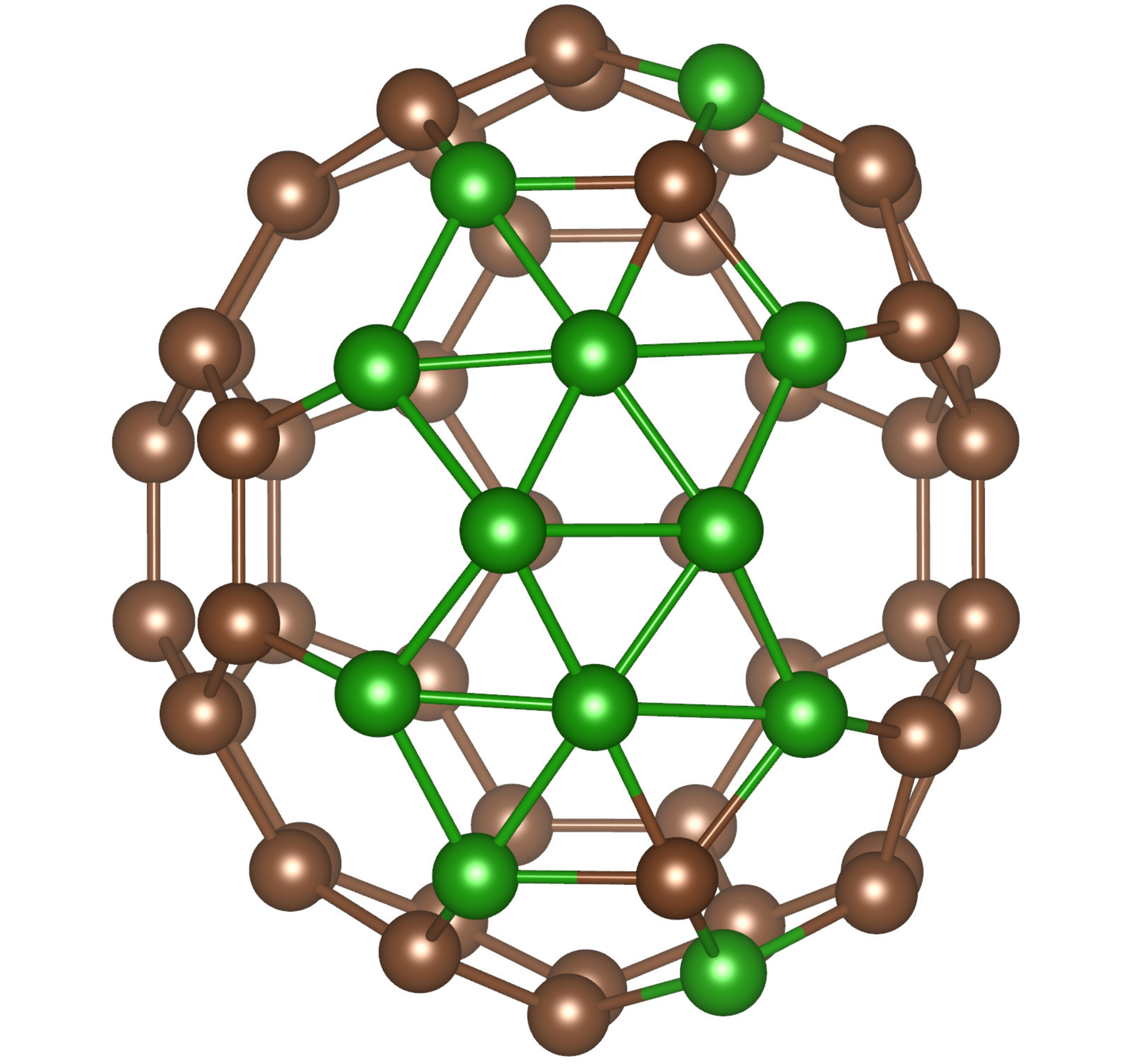}
 \includegraphics[width=0.07\textwidth]{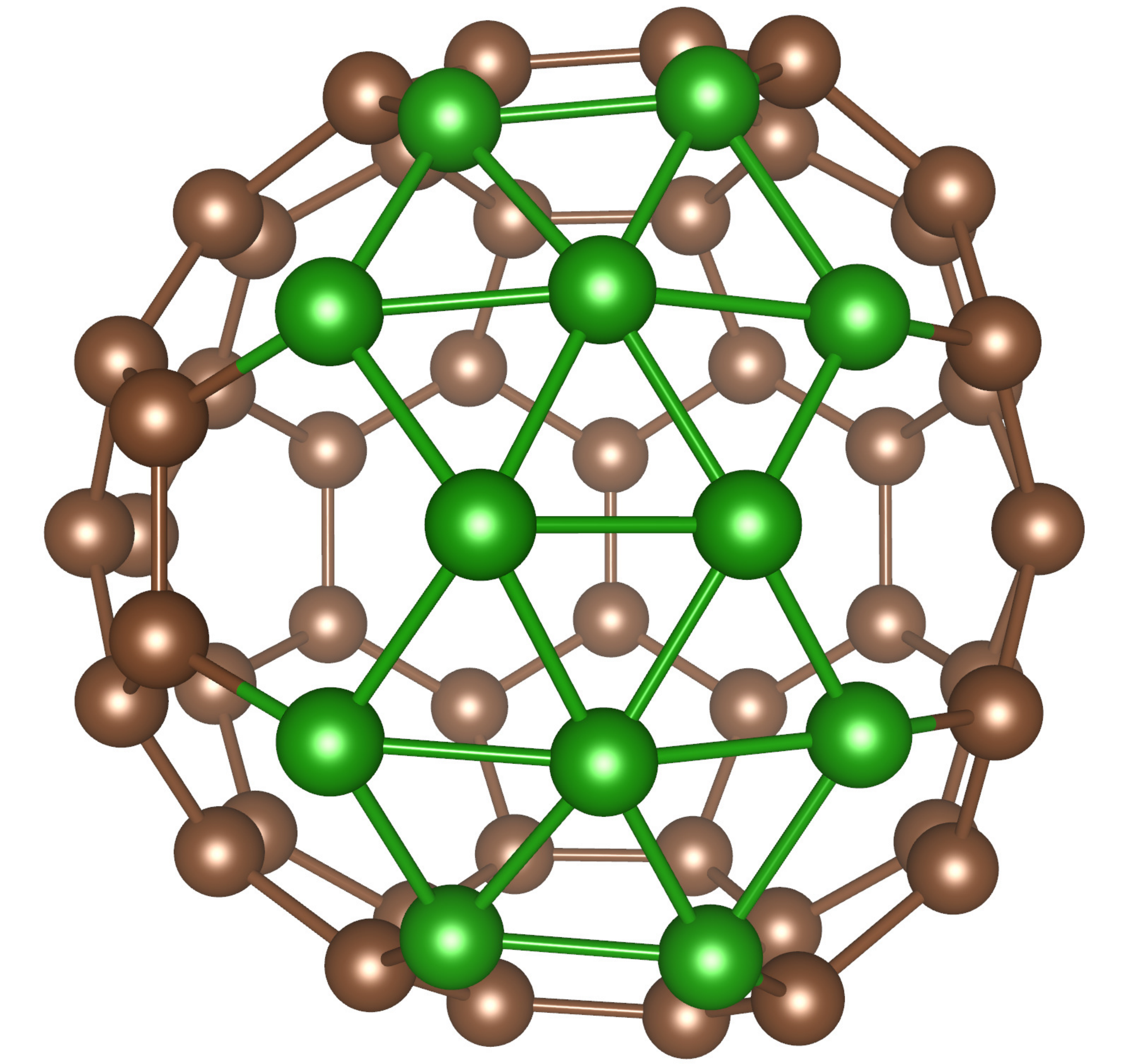}
 \caption{The 12 energetically lowest configurations for the fullerenes \ce{B12C48}.}
 \label{fig:boron-carbon_figures}
\end{figure}

\begin{figure}
 \includegraphics[width=0.1\textwidth]{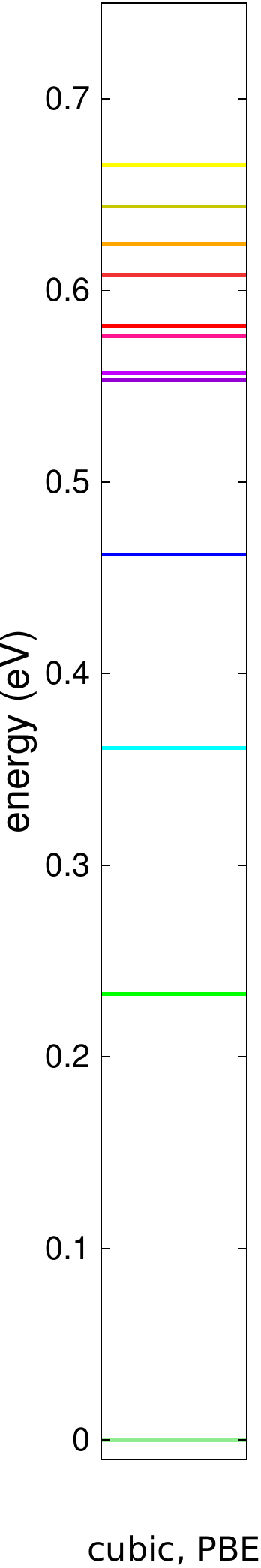}
 \hspace{0.08\textwidth}
 \includegraphics[width=0.1\textwidth]{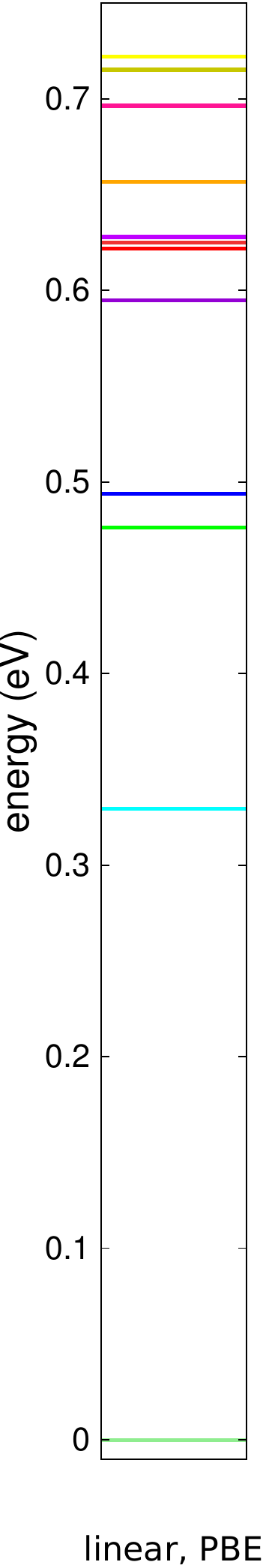}
 \hspace{0.08\textwidth}
 \includegraphics[width=0.1\textwidth]{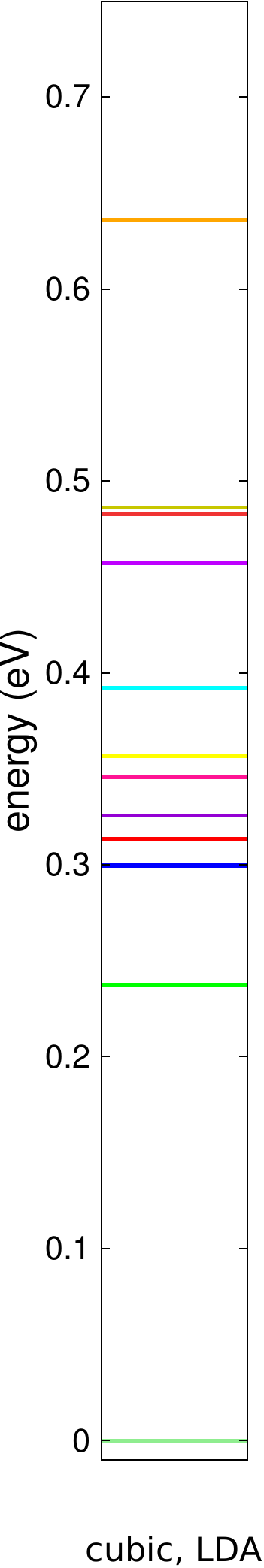}
 \caption{Energetic ordering of the lowest configurations of \ce{B12C48}. From left to right: cubic version with PBE, linear version with PBE, cubic version with LDA. All energy levels have been shifted such that the ground state is always at zero. The colors correspond always to the same level, demonstrating that the energetic ordering is not always the same.}
 \label{fig:boron-carbon_energy_spectrum}
\end{figure}

As an example for a system where the linear scaling version is not able to reproduce the correct results of the cubic version we present the energetic ordering for the 12 energetically lowest structures~\cite{mohr2014boron} of \ce{B12C48}, depicted in Fig.~\ref{fig:boron-carbon_figures}. Such boron-carbon fullerenes are a very delicate system; previous studies~\cite{garg2011substitutional,riadmanaa2003predicted} have even led to different conclusions.
In order to successfully describe the energetic ordering of these configurations it is indispensable to catch energy differences -- i.e.\ signals -- of the order of \unit[1]{meV/atom}, which is possible with the cubic version of BigDFT used in Ref.~\onlinecite{mohr2014boron}.
%For the study of these systems, the cubic BigDFT code has to discriminate energy differences -- ``i.e.\ signals'' -- of the order of \unit[1-4]{meV/atom}.
The energy spectra for these 12 configurations are shown in Fig.~\ref{fig:boron-carbon_energy_spectrum}, for three different setups: the cubic version using the PBE functional, the linear version using as well the PBE functional, and the cubic version using the LDA~\cite{ceperley1980ground} functional. As can be seen, there are -- even if the main features of the system are well captured -- notable differences between  the cubic and the linear version: whereas  the ground state is the same, there has been some reordering of the excited states; in particular the energetic 
gap between the ground state and the first local minimum is wrong. This shows that for this system the linear scaling approach using finite cutoff radii is not appropriate. 
However the third panel demonstrates that these boron-carbon fullerenes are indeed an extremely delicate system. As can be seen, the energetic ordering is also considerably modified if the cubic version is used with another functional: using LDA instead of PBE, the energy levels of the ground state and the first excited one remain identical, but the higher levels are completely jumbled. Due to this very high sensitivity it is thus not surprising that the linear version is not able to satisfactorily reproduce the results of the cubic version, as the noise introduced by the finite cutoff radii is higher than the signal we look for.

\section{Scaling with system size}
\begin{figure}
 \includegraphics[width=.45\textwidth]{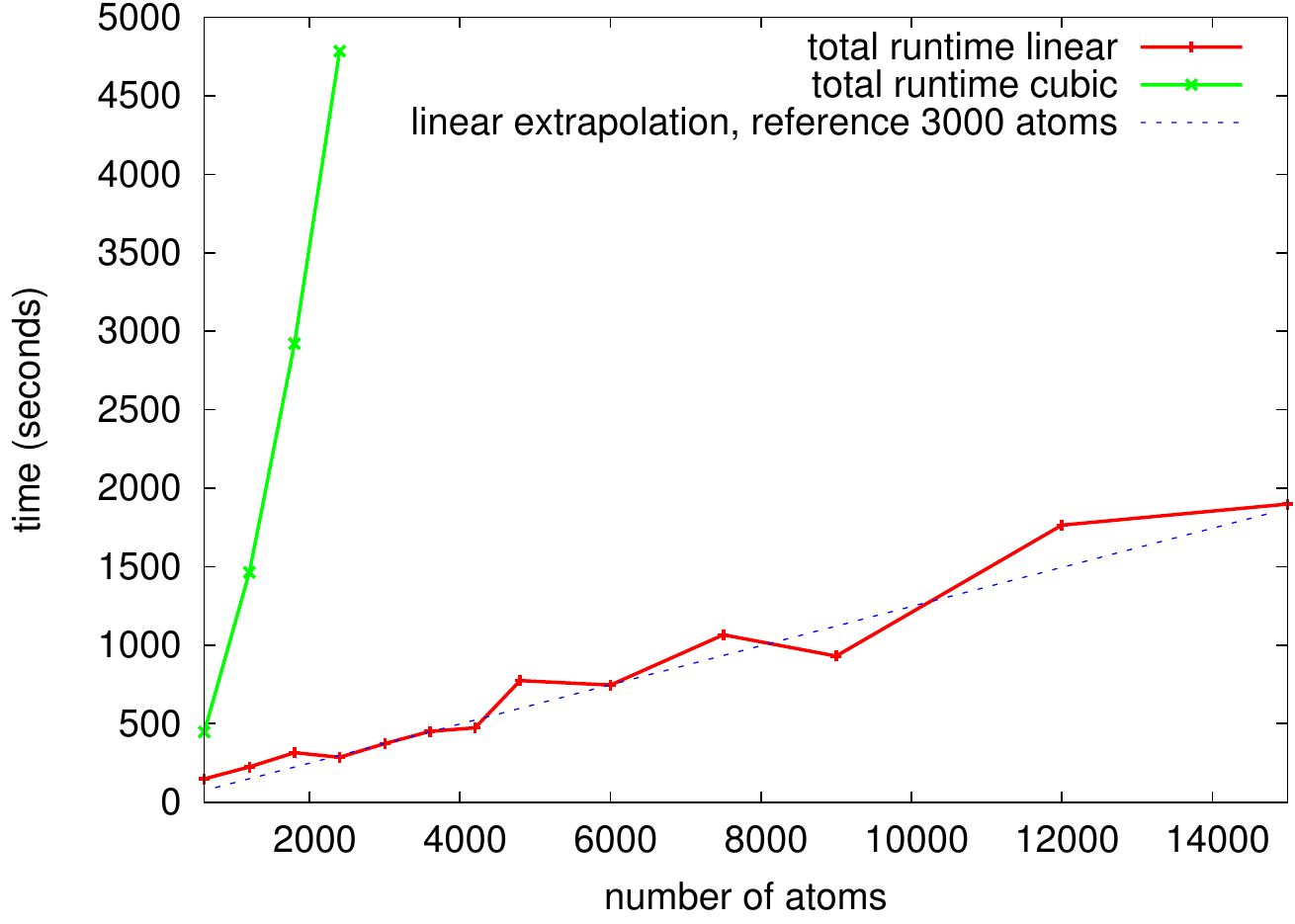}\\
 \vspace{6pt}
 \includegraphics[width=.45\textwidth]{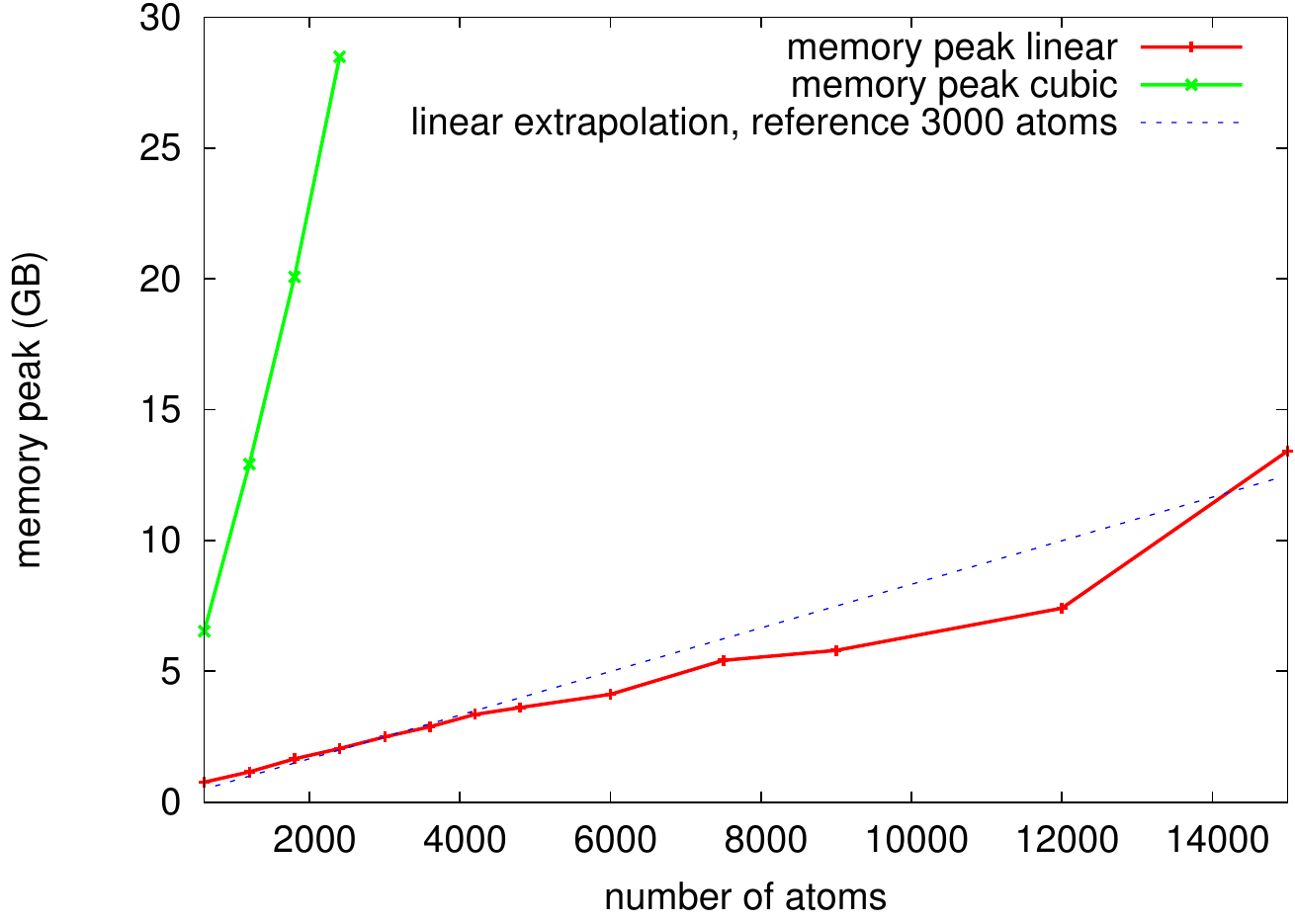}
 \caption{Total runtime for one single point calculation (upper panel) and memory peak per MPI task (lower panel) for water droplets of various size, ranging from 600 to 15000 atoms. The linear scaling version indeed exhibits linear scaling, as indicated by the linear extrapolations. The small deviations are mainly caused by a slightly different number of iterations required to reach convergence.}
 \label{fig:scaling_with_syztemsize_water}
\end{figure}

The goal of this work was to obtain a code which at the same time yields a very high precision and scales linearly. The first property has already been shown in detail in Sec.~\ref{sec:Assessment of the accuracy}; now it remains to demonstrate the second property. To this end we performed single point calculations for randomly generated water droplets of various size, for both the linear and the cubic version. The runs were performed in parallel, using in total 6400 cores. The results for the total runtime and the memory peak are shown in Fig.~\ref{fig:scaling_with_syztemsize_water}. As can be seen both quantities clearly exhibit a strict linear scaling. The cubic version, on the other hand, shows a much steeper increase and does not permit calculations beyond about 2000 atoms. Due to the spherical geometry the degree of sparsity of the matrices is rather low and it is therefore more difficult for the linear scaling version to exploit this property. Nevertheless, even for the smallest droplet the linear scaling 
version is faster than  the cubic one. By extrapolating the curves for the runtime the crossover point for this system can be estimated to be at about 200 atoms, which is in a range still easily accessible by the cubic version of BigDFT for production runs (see e.g.\ Ref.~\onlinecite{Sridevi}).

Moreover it becomes clear that the computational resources consumed by the linear version are only moderate. For a complete single point calculation for 10000 atoms only a few thousand CPU hours are required, which is a rather small amount considering the capacities of current supercomputers. This demonstrates that we were able to drastically reduce the number of degrees of freedom while keeping a very high level of accuracy.
\begin{figure}
 \includegraphics[width=.45\textwidth]{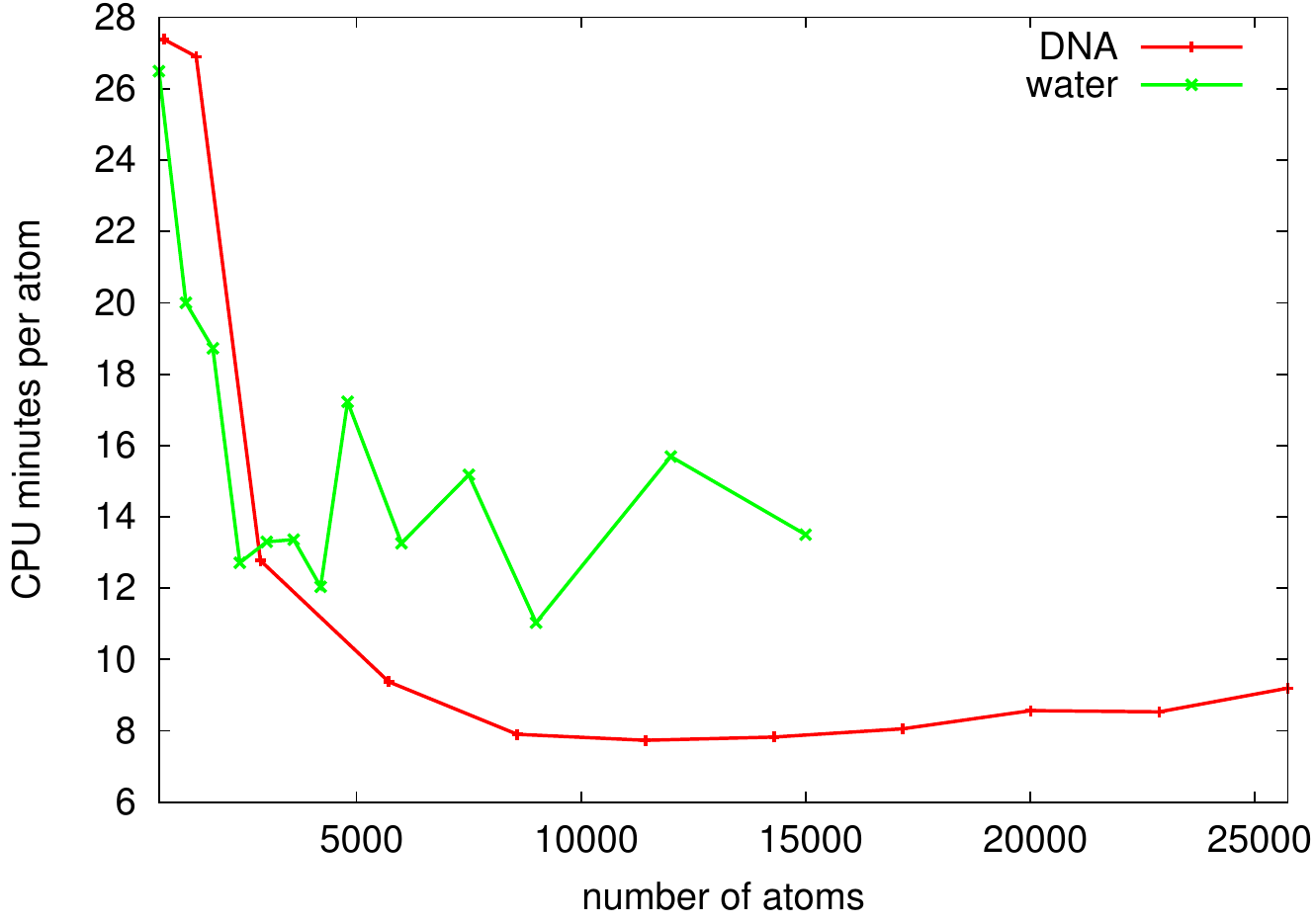}
 \caption{CPU minutes per atom needed for a fully self-consistent calculation for DNA fragments and water droplets of different sizes. As expected, this quantity remains constant above a certain size, with the prefactor mainly influenced by the geometry of the system to be calculated.}
 \label{fig:CPU-minutes_per_atom}
\end{figure}

For a truly linear scaling regime it is also interesting to look at the "CPU minutes per atom" which are needed to perform a fully self-consistent calculation. This quantity is shown in Fig.\ref{fig:CPU-minutes_per_atom} for DNA fragments and the water droplets used for the aforementioned benchmark. As is to be expected for a linear scaling code, this quantity remains constant above a certain critical size where the linear scaling parts of the code start to dominate. Moreover one can easily see how the geometry of the system influences the prefactor, resulting in higher values for the water droplets than for the DNA fragments.

\section{Parallelization}
\subsection{Parallel efficiency}
Even if the linear scaling version of BigDFT requires only moderate resources, it is indispensable to have an efficient parallelization scheme in order to keep the runtimes short and thus to have the possibility of performing advanced calculations such as geometry optimizations within a reasonable time frame. To this end we parallelized our code with a mixed distributed/shared memory parallelization scheme using MPI and OpenMP. It is worth noting that the shared memory parallelization is not just an additional speedup. Reducing the number of MPI tasks and in turn increasing the number of OpenMP threads helps to improve the MPI load balancing and to reduce communication overhead; moreover it can substantially reduce the memory peak per compute node and thus help in situations where this resource is critical.

\begin{figure}
 %\includegraphics[width=0.45\textwidth]{speedup_with_nmpi.pdf} \\
 %\hspace{2pt}
 %\includegraphics[height=0.33\textwidth]{speedup_with_nmpi_fewcores.pdf}
 %\hspace{4pt}
 %\includegraphics[height=0.33\textwidth]{speedup_with_nmpi_manycores.pdf} \\
 \includegraphics[height=0.33\textwidth]{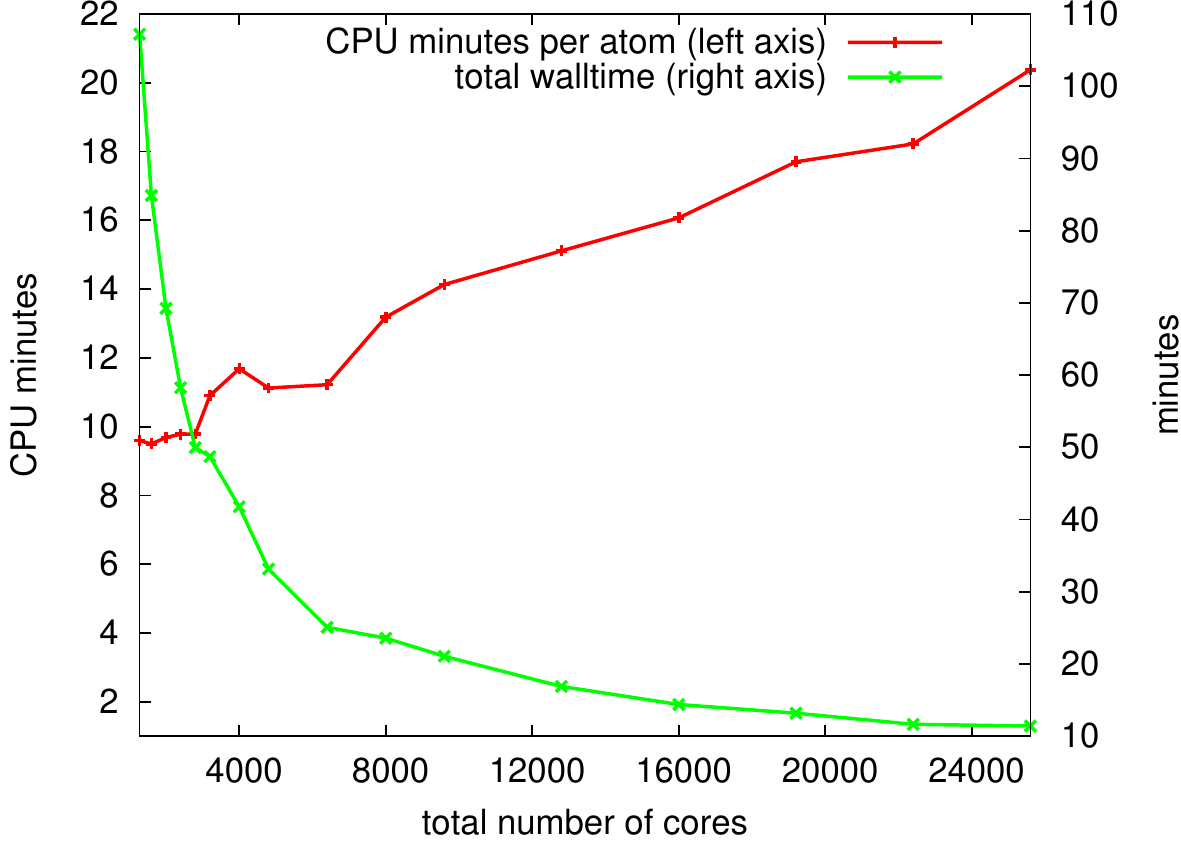} \\
 \vspace{8pt}
 \includegraphics[height=0.33\textwidth]{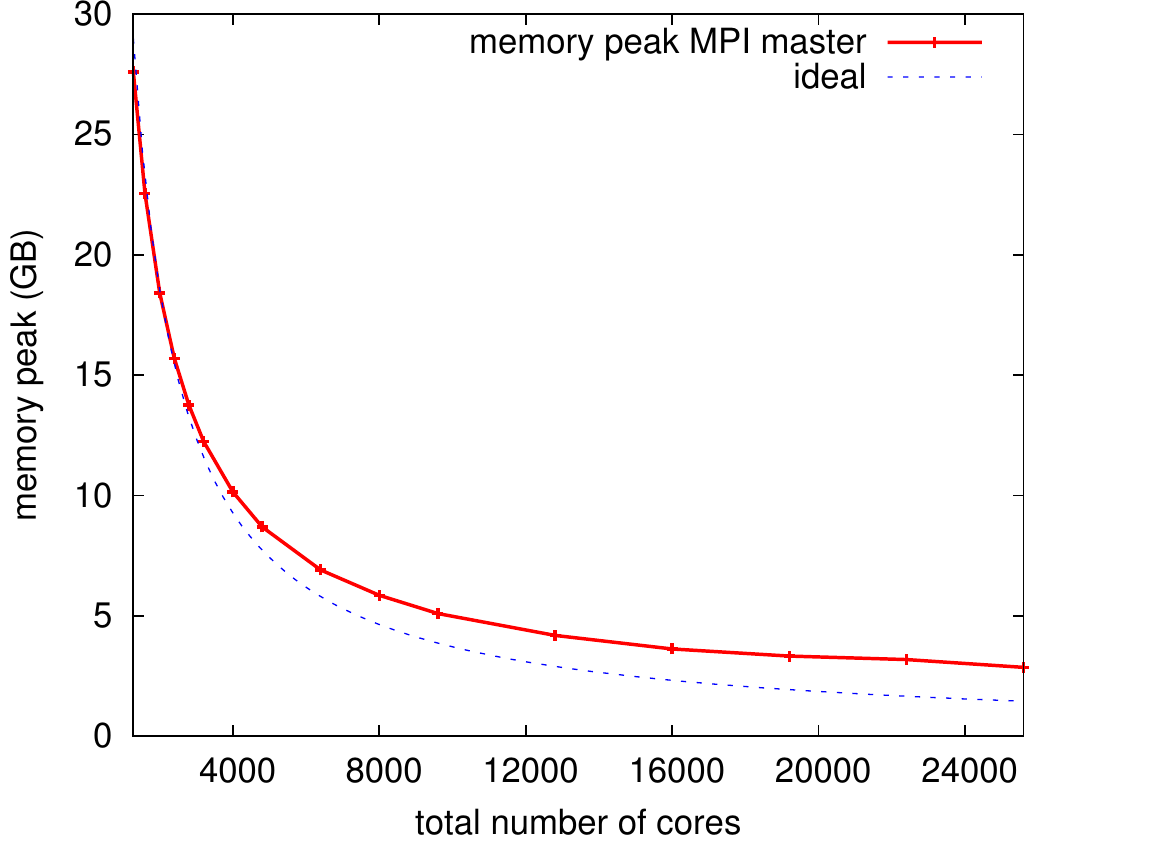}
 \caption{CPU minutes per atom and total walltime (upper panel) and memory peak (lower panel) as a function of the total number of cores for a DNA fragment consisting of 14300 atoms.
 %The plots for the speedup are split up in two ranges: One with the number of cores ranging from 1280 to 6400 (taking as reference 1280 cores), and another one from 6400 to 25600 cores (taking as reference 6400 cores). The runs were all performed using 8 OpenMP threads for each MPI task, thus only the latter number was varied.
 The runs were all performed using 8 OpenMP threads for each MPI task, thus only the latter number was varied.}
 \label{fig:scaling_with_number_of_procs}
\end{figure}

Reaching an efficient parallelization for a linear scaling code is not easy. First of all we have a small number of degrees of freedom and thus little workload than can be shared among the cores. Moreover there are also load balancing problems which can arise due to the different sizes and surroundings of the localization regions. Nevertheless we were able to reach a degree of parallelism which is more than sufficient to efficiently perform large calculations. As an illustration we show the scaling for a fully self-consistent calculation of a DNA fragment consisting of 14300 atoms. The number of MPI tasks ranged from 160 to 3200, each one being split into 8 OpenMP threads; thus in total we have a range of 1280 to 25600 cores. The results for the runtime and the memory peak are shown in Fig.~\ref{fig:scaling_with_number_of_procs}. For the runtime we show the CPU minutes per atom as well as the total walltime. 

As can be seen from the CPU minutes per atoms remaining almost constant the speedup is very good up to a few thousand cores. 
%-- in particular in view of the fact that we use already 1280 cores as the first data point. 
On the other hand this quantity increases for very large number of cores, revealing that the speedup becomes only moderate in this range. However this does not mean that the code is poorly parallelized. The reason is rather that, despite of the considerable size of the system, the number of degrees of freedom is so small that using too many cores simply results in a very small number of operations to be executed by a single core and thus to a poor ratio of computation and communication. Moreover it is more difficult to reach an efficient load balancing in such a situation, resulting in many cores being idle most of the time. Consequently this degradation in the parallel speedup will be shifted towards larger values if one increases the system size. 

Nevertheless it is worth noting that even for the largest number of cores there is still some speedup, as can be seen from the total walltime decreasing steadily.
Even in this range, the CPU time per atom remains of the same order as the one needed at the crossover point with the cubic code. As calculations with the cubic BigDFT code are already accessible in this range (as pointed out in the previous section), production runs of very large systems might become feasible by linearly scaling the computing resources needed at the crossover point.
The memory parallelization, shown in the lower panel, does not suffer from any degradation, and we come close to a perfect scaling. In summary these results show that our code has an excellent level of parallelism as long as one keeps a good balance between the size of the system and the computational resources that are utilized.

\subsection{Parallelization of the sparse matrices}
Some strategies to obtain an efficient parallelization have already been outlined in Ref.~\onlinecite{mohr2014daubechies}. Here we present a new concept which we developed to reach an efficient handling of the sparse matrices.

The problem of the transposed approach~\cite{mohr2014daubechies} for calculating the overlap matrix (and similarly the Hamiltonian matrix) is that it requires a global reduction operation at the end. This can pose a bottleneck both from the viewpoint of runtime and memory. Moreover this global reduction is wasteful due to the locality which is inherent to the linear scaling approach. Each MPI task only needs to know a small portion of the entire matrices and a global reduction is thus not necessary. To circumvent this issue, the MPI tasks are regrouped in taskgroups, sharing a portion of the entire matrix -- obviously this portion is chosen such that it contains those parts of the global matrix which are actually needed by each individual task. In addition the taskgroups are defined such that each MPI task belongs to at most two taskgroups. This has two advantages: first of all each MPI task only needs to hold a copy of a part of the global matrix, thus reducing the memory requirements, and secondly the reduction only needs to be performed within a taskgroup. Thanks to the use of non-blocking collective MPI routines a task can participate in two reduction operations at the same time without the risk of serializing the code.

\begin{figure}
 \includegraphics[width=0.45\textwidth]{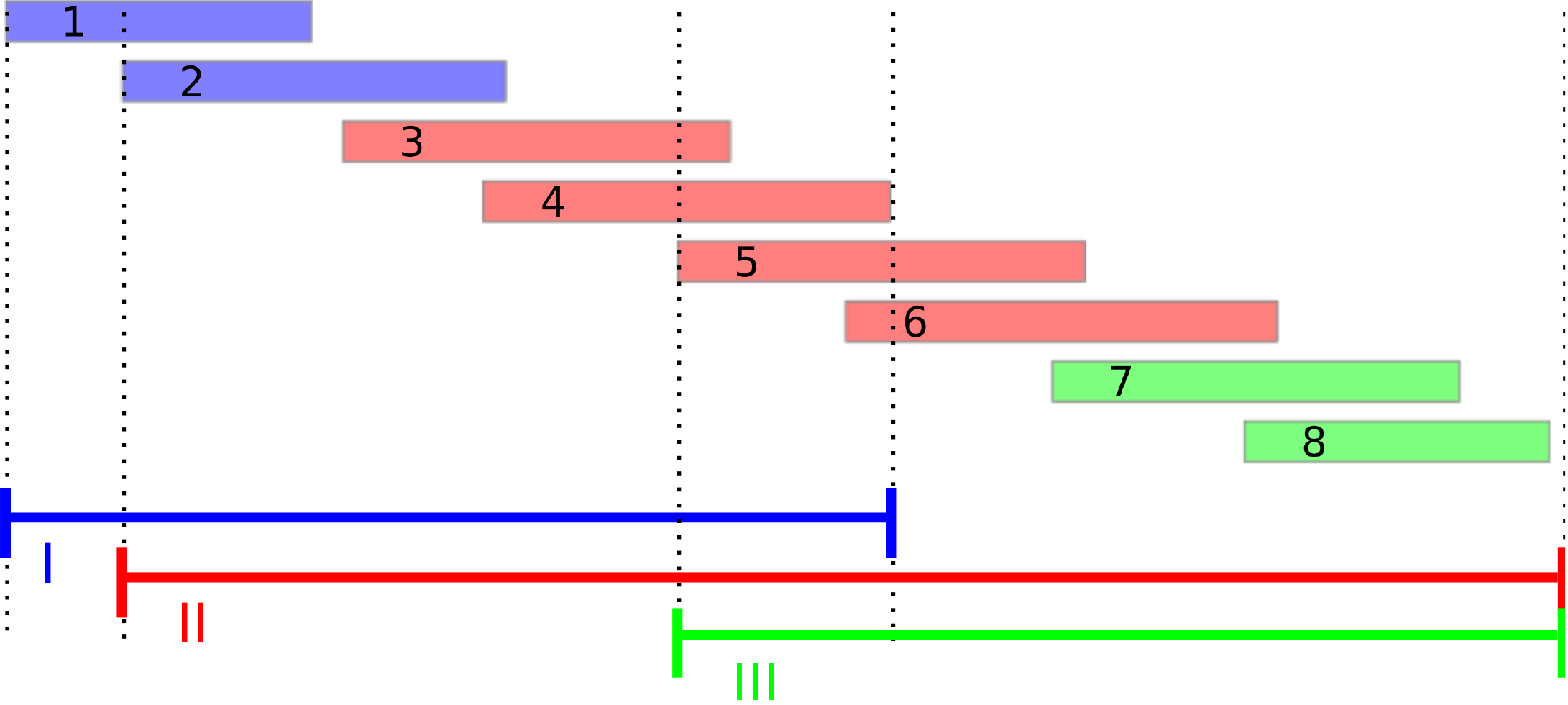}
 \caption{Illustration of the concept of the matrix taskgroups for a toy system. There are 8 MPI tasks which are regrouped in 3 taskgroups. Each taskgroup is initially defined such as to form a matrix subset, and subsequently extended to include all the processes which share data of the initial subset. Care is taken in not including each MPI task in more than two different taskgroups.}
 \label{fig:matrix_taskgroups}
\end{figure}

The concept is visualized in Fig.~\ref{fig:matrix_taskgroups} for a toy system consisting of 8 MPI tasks. The matrix subparts which are needed by each MPI task are indicated by rectangles. The MPI tasks are initially regrouped in 3 taskgroups: taskgroup I is responsible for the part needed by the tasks 1-2, taskgroup II for those needed by the tasks 3-6, and taskgroup III for those needed by the tasks 7-8. The taskgroups are then enlarged in order to guarantee that the reduction is done correctly. Therefore taskgroup I includes the tasks 1-4, taskgroups II includes the tasks 2-8, and taskgroup III includes the tasks 5-8.\\
Obviously the toy system is too small to get a real benefit from the taskgroups. However for large systems there can be a substantial gain. As an example we show in Tab.~\ref{tab:matrix_taskgroups_timings} the timings obtained with and without the matrix taskgroups for two specific operations. As can be seen, the calculation of the overlap and Hamiltonian matrix (both computation and communication) can be accelerated by more than a factor of 3, demonstrating that actually most of the time was spent in the reduction and not in the computation itself. The other operation, which is required at the end of the FOE procedure to build up the density kernel out of the partial results calculated by each task and is almost entirely communication based, can even benefit from a speedup of 16. 
\begin{table}
 \scriptsize
 \begin{tabular}{l c c}
  \hline\hline
         & \multicolumn{1}{c}{overlap calculation}       & \multicolumn{1}{c}{gather/compress} \\
         & \multicolumn{1}{c}{\scriptsize seconds} & \multicolumn{1}{c}{\scriptsize seconds} \\
           \hline
  without taskgroups & 88.9                  & 266.9 \\
  with taskgroups    & 25.2                  & \phantom{2}16.4  \\ %phantom to make right-aligned
  \hline\hline
 \end{tabular}
 \caption{Speedups offered by the matrix taskgroups for two specific operations.}
 \label{tab:matrix_taskgroups_timings}
\end{table}

\section{Conclusion}
Using a set of strictly localized and quasi-orthogonal support functions which are adapted in-situ to their chemical environment we were able to develop a code which can perform accurate DFT calculations with a strict linear scaling with respect to the size of the system. Thanks to the compact support property of the underlying Daubechies wavelets, this reduction of the complexity with respect to the traditional cubic approach does not come at the cost of a loss of precision. Indeed, we are able to get an excellent level of accuracy with a set of standard input parameters and can thus get rid of the tedious fine tuning which is often needed for other $\mathcal{O}(N)$ approaches. We believe this would considerably simplify the usage of this code by end-users, as this fine-tuning usually restricts the usage $\mathcal O(N)$ approaches to a community of specialists.

Moreover we were able to drastically cut down the degrees of freedom needed for an accurate calculation, thus reducing the prefactor of the scaling law and consequently considerably diminishing the amount of computational resources required. Even for systems containing 10000 atoms a complete single point calculation can be done using only of the order of one thousand CPU hours. 
Furthermore the linear scaling behavior of the program motivates the notion of ``CPU minutes per atom'', making it easy to estimate the computational resources needed for the simulation of a given system of any size.
Together with the efficient parallelization scheme this low consumption of computational resources also paves the way towards more expensive tasks such as geometry optimizations.

\textbf{Acknowledgments.} S. M. acknowledges a postdoctoral fellowship from the CEA Nanosciences horizontal aid programme (project BigPOL).  T.D and L. G. are indebted to the French National Research Agency for funding under Contract No. ANR-2010-COSI-005-01 NEWCASTLE. Computing time has been provided by the "Curie" national GENCI-IDRIS supercomputing center under contract No. i2014096905. Computing time of the CSCS Swiss project s499 is also acknowledged.

\bibliography{citationlist}

\end{document}